\documentclass[12pt,preprint]{aastex}
\usepackage{epsf}

\newcommand{\be}{\begin{displaymath}}
\newcommand{\ee}{\end{displaymath}}
\newcommand{\bea}{\begin{eqnarray}}
\newcommand{\eea}{\end{eqnarray}}

\shortauthors{Pavel Denissenkov}
\shorttitle{Thermohaline Convection in RGB Stars}

\begin{document}

\title{NUMERICAL SIMULATIONS OF THERMOHALINE CONVECTION: IMPLICATIONS FOR EXTRA-MIXING
       IN LOW-MASS RGB STARS}

\author{Pavel A. Denissenkov}
\affil{Department of Physics \& Astronomy, University of Victoria,
       P.O.~Box 3055, Victoria, B.C., V8W~3P6, Canada}
              \email{pavel.denisenkov@gmail.com}
 
\begin{abstract}

Low-mass stars are known to experience extra-mixing in their radiative zones on the red-giant branch (RGB) above the bump luminosity.
To determine if the salt-fingering transport of chemical composition driven by $^3$He burning is efficient enough to produce RGB extra-mixing,
2D numerical simulations of thermohaline convection for physical conditions corresponding to the RGB case
have been carried out. We have found that
the effective ratio of a salt-finger's length to its diameter $a_{\rm eff}\la 0.5$ is more than ten times smaller than the value
needed to reproduce observations ($a_{\rm obs}\ga 7$). On the other hand, using the thermohaline diffusion coefficient from
linear stability analysis together with $a=a_{\rm obs}$ is able to
describe the RGB extra-mixing at all metallicities so well that it is tempting to believe that it may represent
the true mechanism. In view of these results, follow-up 3D numerical simulations of
thermohaline convection for the RGB case are clearly needed.

\end{abstract} 


\section{Introduction}

A main sequence (MS) star with a mass $M\la 1.5\,M_\odot$, e.g. the Sun, gets its energy from
thermonuclear fusion reactions of the pp-chains. When its core has exhausted the hydrogen fuel,
the star leaves the MS and becomes a red giant branch (RGB) star. The red giant has a compact electron-degenerate
helium core of the size of the Earth surrounded by a thin hydrogen burning shell (hereafter, HBS) that provides
the star with the energy via the transformation of H into He via the CNO-cycle. The bulk of the red giant's volume
is occupied by a convective envelope that extends from the surface to nearly the HBS, however the base of
the convective envelope always stays separated from the HBS by a convectively stable radiative zone
with a thickness of the order of one solar radius. 

It is possible to select a large sample of low-mass post-MS stars
that have very similar masses and metallicities. The simplest way to do this is to choose normal post-MS stars belonging
to a same several billion year old open or globular cluster. For field stars, one needs well-measured distances to
estimate the masses accurately enough, so that only stars with well-determined Hipparcos parallaxes suit this purpose.
A distribution of the characteristics of stars from such a sample along the RGB in the Hertzsprung-Russell (HR)
diagram can then be interpreted as an evolutionary
sequence (the Vogt-Russell theorem). In particular, a correlation of a surface chemical composition anomaly
(a deviation of atmospheric abundances from those on the MS) with the luminosity or effective temperature
can be considered as an evolutionary pattern produced by nuclear reactions and ongoing mixing.

Spectroscopic observations and their theoretical interpretations have firmly established two distinct mixing episodes in low-mass RGB stars (e.g., \citealt{ch94}).
First, on the lower RGB, the convective envelope quickly grows in mass and, as a result, its base penetrates the layers in which the partial
burning of H via the pp-chains resulted in an accumulation of large amounts of $^3$He, and where Li was mostly destroyed. It also
reaches the layers in which the non-equilibrium operation of the CN-branch of the CNO-cycle took place.
This should lead to sharp changes of the star's atmospheric chemical composition, such as a strong decrease of
its surface Li abundance, a noticeable reduction of the $^{12}$C/$^{13}$C isotopic ratio, a modest
decrease and increase of the C and N abundances respectively, and a considerable enrichment of the convective envelope in $^3$He.
These changes continue until the convective envelope attains its maximum mass after which its base
starts to recede in front of the HBS advancing in mass. This mixing episode is called the first dredge-up. All of the above predicted changes of the star's
surface chemical composition, except those of $^3$He, have often been observed on the lower RGB.

Second, on the upper RGB, above the luminosity at which differential luminosity functions of globular-cluster RGB stars
have clearly visible bumps, observations reveal that the same abundance changes that
were occurring during the first dredge-up resume, and that the changes are even larger this time.
The bump luminosity corresponds to the evolutionary phase during which the HBS passes through and erases the chemical composition
discontinuity imprinted by the convective envelope at the end of the first dredge-up.  When the HBS crosses
the discontinuity it finds itself in a region with a slightly increased H abundance; consequently, the red giant readjusts its structure appropriately,
causing it to move to slightly lower luminosities before continuing up the RGB.
As a result of spending more time in the narrow range of luminosity where this readjustment occurs,
there is a pile-up of stars around the bump luminosity. Given that the base of convective envelope recedes
through a region of homogeneous chemical composition, the observed evolutionary variations of the surface abundances
on the upper RGB can be understood only if there is some extra-mixing process of non-convective origin that
connects the convective envelope with the HBS.
The physical nature of this extra-mixing still remains elusive. 
For a long time, it has been thought that the RGB extra-mixing is related to
rotation-induced instabilities (e.g., \citealt{sm79,ch95,dt00}). However, the work of \cite{pea06} casts doubt on these models.

It was not until recently that the most promising mechanism of the RGB extra-mixing was proposed by \cite{eea06} and
\cite{chz07}. When the HBS is approached from the base of convective envelope, 
the first active thermonuclear reaction involving abundant nuclei that is encountered     
is $^3$He($^3$He,2p)$^4$He. It has a unique property of lowering the mean molecular weight locally, by $\Delta\mu\approx\mu^2\Delta X(^3\mbox{He})/6$,
where $\Delta X(^3\mbox{He})$ is a mass fraction of $^3$He consumed in the reaction.
As a consequence, the density is reduced by the same amount (assuming the ideal gas law), which makes the local material lighter
than that in its immediate surroundings, therefore the $^3$He-depleted material will tend to rise.
Mixing driven by $^3$He burning in stars was discussed already in the 1970s (e.g., \citealt{u72}), however it was the publication by
\cite{eea06} that brought this mechanism to the attention of other researchers in the context of the RGB extra-mixing.

\cite{chz07} have given a correct physical explanation of the mixing process to which the $^3$He burning in the HBS
should lead. This is the salt-fingering or thermohaline convection that has been studied in detail, both experimentally
and theoretically, in oceanography for many years (e.g., see reviews by \citealt{rk03}, \citealt{sch03}, and \citealt{k03}).
It is caused by a double-diffusive instability that occurs in a situation when a stabilizing agent (heat) diffuses away
faster than a destabilizing agent (salt and $\mu$ in the oceanic and RGB cases, respectively). In the ocean,
the double-diffusive instability usually develops close to the surface where warm salty water finds itself
overlying cold fresh water. A blob of the surface water will tend to sink because its higher salinity (a concentration of salt)
makes it denser than the surrounding deeper water while its temperature remains close to the ambient one thanks to
the faster diffusion of heat. Similarly, a blob at depth will tend to rise. In the oceanic case, the usual outcome of
this instability are vertically elongated salt fingers containing sinking and rising water of different salinity.

Given that the local reduction of $\mu$ by the $^3$He burning is very small, $\Delta\mu/\mu\sim -10^{-4}$, it can become visible
only in the background of homogeneous chemical composition. This happens precisely at the bump luminosity. Another feature that
makes the $^3$He-driven thermohaline convection the most promising mechanism for the RGB extra-mixing is its dependence on just one
parameter, the envelope $^3$He abundance after the first dredge-up. This means that stars with similar masses and metallicities
should demonstrate comparable chemical composition anomalies on the upper RGB, the pattern that seems to be observed in real stars
(e.g., \citealt{grea00,sm03}). Before the RGB extra-mixing was introduced into the standard stellar evolution theory, the latter
had been in conflict with the seemingly observed constancy of the interstellar $^3$He abundance since the Big Bang.
The problem is that low-mass stars produce a lot of $^3$He on the MS that is dredged up to the surface on the RGB and
then deposited in the interstellar medium as a result of mass-loss. The most plausible solution of this problem is
the RGB extra-mixing that strongly decreases the envelope abundance of $^3$He by circulating it through the $^3$He burning layers
of the HBS. Therefore, if the RGB extra-mixing is indeed driven by the $^3$He burning (thermohaline convection) then the cosmological
$^3$He problem has a beautiful solution: it is $^3$He itself abundantly produced in low-mass MS stars that
takes care of its own destruction in the same stars on the RGB, so that the net balance of $^3$He from low-mass stars in the interstellar medium is close to zero.

\cite{chz07} have modeled thermohaline mixing using a diffusion coefficient obtained from a linear theory by \cite{u72}. Unfortunately, the linear theory
does not give a reliable estimate of the maximum length of salt fingers relative to their diameter, i.e., the finger aspect ratio $a=l/d$,
the square of which enters the diffusion coefficient. This leads to a large uncertainty in the theory leaving it basically semi-empirical,
like other up-to-date theories and models of the RGB extra-mixing. From the results presented by \cite{chz07}, it follows that the surface abundance
patterns in low-mass RGB stars can be reproduced theoretically only if $a\gg 1$. To support the large finger aspect ratio,
\cite{chz07} referred to the oceanic case where long salt fingers were observed experimentally. However, it is not legitimate to directly
compare the oceanic and RGB cases because they correspond to very different fluid flows sets. In particular, the ratio of viscosity to
thermal diffusivity (the Prandtl number $Pr$) is close to ten for the oceanic case, whereas $Pr < 10^{-5}$ for the RGB case.
It turns out that only direct numerical simulations with non-linear interactions and other relevant effects 
taken into account can give better insight into the $^3$He-driven 
thermohaline mixing in the low-mass RGB stars. It is this task that is addressed in this research for the first time.

The paper is organized as follows. Section \ref{sec:linear} summarizes the main results from the linear theory that are relevant
for our further discussion. Section \ref{sec:numeric}
presents and analyzes the results of our 2D numerical simulations of thermohaline convection. It is followed by Section \ref{sec:1d}, in which 
we employ a diffusion coefficient from the linear theory to
model, as precisely as possible, the changes of chemical composition incurred by the RGB thermohaline mixing.
Discussion and conclusions are provided in the final Section \ref{sec:concl}.
Wherever it is appropriate, we make a comparison of the oceanic and RGB cases.

\section{Relevant Results from the Linear Salt-Fingering Theory}
\label{sec:linear}

In this as well as in the next section, our analysis begins with the Boussinesq equations that describe motion in a nearly
incompressible stratified viscous fluid
\bea
\label{eq:bouss1}
\frac{\partial\mathbf{v}}{\partial t} + (\mathbf{v},\nabla)\mathbf{v} & = & \frac{\delta\rho}{\rho_0}\,\mathbf{g} + \nu\nabla^2\mathbf{v}, \\
\label{eq:bouss2}
\frac{\partial T}{\partial t} + (\mathbf{v},\nabla)T & = & k_T\nabla^2T, \\
\label{eq:bouss3}
\frac{\partial S}{\partial t} + (\mathbf{v},\nabla)S & = & k_S\nabla^2S,
\eea
where $\mathbf{v}$ is the velocity, $\rho_0$ is a constant reference density, $\delta\rho$ is a deviation of the local density from $\rho_0$,
$\mathbf{g}$ is the gravitational acceleration, $\nu$ is the viscosity, $T$ is the temperature, $S$ is the salinity, $k_T$ and $k_S$ are the thermal and haline  
diffusivities. Note that, in the RGB case, $S$ should be replaced by $\mu$, while equation (\ref{eq:bouss3}) is still valid as long as
$|\Delta\mu|/\mu\ll 1$. We have also neglected source terms in equations (\ref{eq:bouss2}) and (\ref{eq:bouss3}) that could be related to nuclear reactions
in the RGB case. 

In the Boussinesq approximation, which is a reasonable one even for a compressible fluid provided that
its motion is studied on length scales much less than the density height scale and velocities remain much less than the speed of sound, 
the density variation is taken into account only in the buoyancy force term $\mathbf{f}_{\rm b} = (\delta\rho/\rho_0)\,\mathbf{g}$.
Assuming that the relative deviations of $\rho$, $T$, and $S$ from their reference values in the initial unperturbed state $(\rho_0,T_0,S_0)$
are small, we use a linearized equation of state
\bea
\delta\rho = -\alpha\delta T + \beta\delta S,
\label{eq:eos}
\eea
where 
$$
\alpha = -\frac{1}{T}\left(\frac{\partial\ln\rho}{\partial\ln T}\right)_P,\ \ \ \mbox{and}\ \ \ 
\beta = \frac{1}{S}\left(\frac{\partial\ln\rho}{\partial\ln S}\right)_P
$$
are the coefficients of thermal expansion and haline contraction. For the ideal gas law, which provides a good approximation
to the equation of state in the radiative zone of a low-mass RGB star, we simply have $\alpha = 1/T$ and $\beta = 1/\mu$.

We use the Cartesian coordinate system $(x,y,z)$ oriented so that its vertical axis $z$ has a direction
opposite to that of the gravitational acceleration and the $x$-axis is located in a star's meridional plane. 
Let $w$ denote the velocity's vertical component. Linearizing equations (\ref{eq:bouss1}\,--\,\ref{eq:bouss3}),
we arrive at the following system of linear PDEs for the vertical velocity component and variations of temperature and salinity:
\bea
\label{eq:lin1}
\frac{\partial w}{\partial t} & = & g(\alpha\delta T - \beta\delta S) + \nu\nabla^2 w, \\
\label{eq:lin2}
\frac{\partial\delta T}{\partial t} & = & -w\frac{\partial T}{\partial z} + k_T\nabla^2\delta T, \\
\label{eq:lin3}
\frac{\partial\delta S}{\partial t} & = & -w\frac{\partial S}{\partial z} + k_S\nabla^2\delta S.
\eea
Given that we seek a solution representing vertically elongated structures with a large ratio of the vertical to horizontal
length scales, we can neglect horizontal velocity components
and we can also assume that $\nabla^2\approx \partial^2/\partial^2 x + \partial^2/\partial^2 y$.

Following \cite{k87}, we will take into consideration the influence of
salt-fingering on the ambient temperature and salinity gradients that appear on the right-hand sides of
equations (\ref{eq:lin2}) and (\ref{eq:lin3}). This influence is determined by the ratios $\delta_T = -2\delta T/\Delta T$ and
$\delta_S = -2\delta S/\Delta S$, in which $\Delta T = (\partial T_0/\partial z)\,l$ and $\Delta S = (\partial S_0/\partial z)\,l$
are differences in temperature and salinity between two points separated by the finger length $l$ in the vertical direction
in the initial unperturbed state. The factor of two in the above relations takes account of the fact that every sinking finger
with excesses of S and T has a neighbouring rising finger possessing deficiencies of S and T of same magnitudes.
The perturbed gradients are expressed via the new independent variables
\bea
\label{eq:gradt}
\frac{\partial T}{\partial z} = \frac{\partial T_0}{\partial z}\,(1-\delta_T), \\
\label{eq:grads}
\frac{\partial S}{\partial z} = \frac{\partial S_0}{\partial z}\,(1-\delta_S),
\eea
where $0\leq(\delta_T,\,\delta_S)\leq 1$.
With regard to the temperature gradient, it is important to note a difference between the oceanic and RGB cases.
In the RGB case, one should subtract the adiabatic temperature gradient $(\partial T/\partial z)_{\rm ad}$ from both $\partial T/\partial z$ and
$\partial T_0/\partial z$ in equation (\ref{eq:gradt}) because a temperature gradient is always negative in stars and it remains stable only as long as
its absolute magnitude is less than that of the adiabatic gradient. The maximum effect that a salt-fingering heat flux
can produce on $\partial T/\partial z$ is to render it adiabatic ($\delta_T = 1$). In the oceanic case, any positive temperature gradient stabilizes
the density stratification. 

Finally, expressing $\delta T$ and $\delta S$ by $\delta_T$ and $\delta_S$ and substituting
them into the system (\ref{eq:lin1}\,--\,\ref{eq:lin3}) gives
\bea
\label{eq:linfin1}
\frac{\partial w}{\partial t} - \nu\nabla^2 w & = & \frac{l}{2}\,g\beta\,\frac{\partial S_0}{\partial z}\,(\delta_S - R_\rho\delta_T), \\
\label{eq:linfin2}
\frac{\partial\delta_T}{\partial t} - k_T\nabla^2\delta_T & = & \frac{2}{l}\,w\,(1-\delta_T), \\
\label{eq:linfin3}
\frac{\partial\delta_S}{\partial t} - k_S\nabla^2\delta_S & = & \frac{2}{l}\,w\,(1-\delta_S),
\eea
where $R_\rho = \alpha\Delta T/\beta\Delta S$ is a parameter known as the density ratio in oceanography.
In stellar physics, it corresponds to $R_\rho = (\nabla - \nabla_{\rm ad})/\nabla_\mu$, where $\nabla = (\partial\ln T_0/\partial\ln P)$,
$\nabla_{\rm ad} = (\partial\ln T/\partial\ln P)_{\rm ad}$, and $\nabla_\mu = (\partial\ln\mu_0/\partial\ln P)$.
Solutions of the last equations are sought in the usual form of $w$, $\delta_T$, $\delta_S \propto \exp(\sigma t)\times\exp[\,i(k_x x + k_y y)]$
taking into account that the vertical velocity can be approximated as $w\approx\sigma l/2$, where $\sigma$ is the growth rate of
salt fingers, whereas $k_x$ and $k_y$ are their horizontal wave numbers. After some simple algebra, the three equations are reduced
to the following third-order dispersion relationship:
\bea
\nonumber
4\sigma^3 + 2(k_T+k_S+2\nu)k^2\sigma^2 + \{[k_Tk_S+2\nu(k_T+k_S)]k^4 + \\
2g\beta\,\frac{\partial S_0}{\partial z}\,(R_\rho-1)\}\sigma -
g\beta\,\frac{\partial S_0}{\partial z}\,(k_T-R_\rho k_S)k^2 = 0,
\label{eq:dispers}
\eea
where $k = \sqrt{k_x^2+k_y^2}$ is related to the salt-finger diameter $d = 2\pi/k$.

\begin{figure}
\epsffile [60 260 580 695] {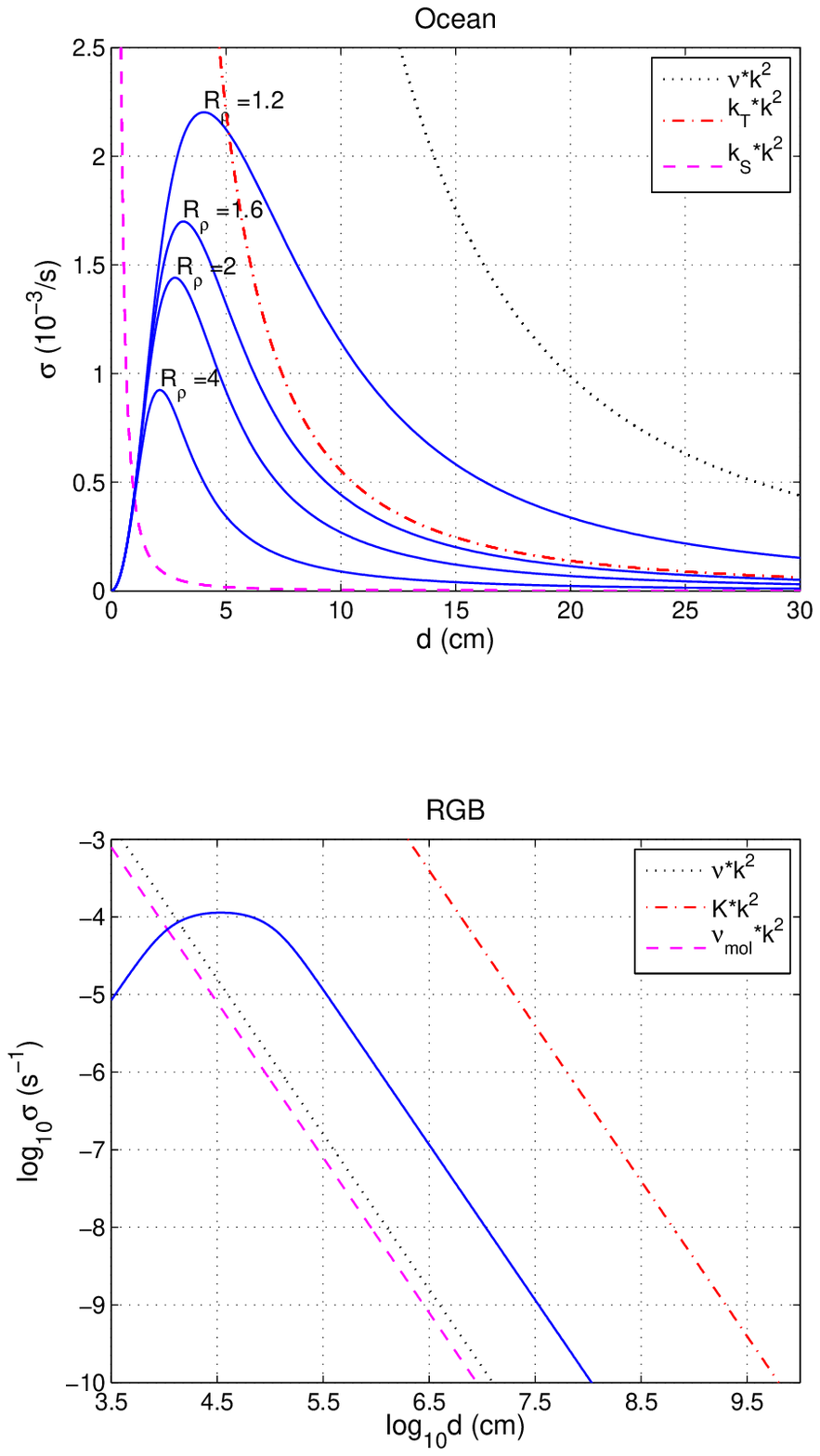}
\caption{Solid curves show growth rates for salt fingers of different diameters for the oceanic (upper panel) and RGB (lower panel) cases.
         Dotted, dot-dashed and dashed curves are reciprocals of microscopic viscous, thermal and mixing timescales for perturbations with
         the wave number $k=2\pi/d$.
         }
\label{fig:f1}
\end{figure}

Positive real roots of equation (\ref{eq:dispers}), corresponding to exponentially growing solutions
(the double-diffusive instability), are plotted in Fig.~\ref{fig:f1} for the oceanic (upper panel)
and RGB (lower panel) cases. Note that, from here on, we will be using more customary notations for
the stellar structure parameters related to salt-fingering, whereas those introduced so far will be reserved
for the oceanic case. In particular, the thermal and haline diffusivities become
the radiative and molecular diffusivities in the RGB case
\bea
\nonumber
k_T & \rightarrow & K = \frac{4acT^3}{3\varkappa C_P\rho^2}, \\ \nonumber
k_S & \rightarrow & \nu_{\rm mol} = 1.84\times 10^{-17}(1+7X)\,\frac{T^{5/2}}{\rho}\ (\mbox{cm}^2\,\mbox{s}^{-1}),
\eea
where $a$ is the radiation constant, $c$ is the speed of light, $\varkappa$ is the Rosseland mean opacity,
$C_P$ is the specific heat at constant pressure, and $X$ is the hydrogen mass fraction.
The viscosity in the vicinity of the HBS consists of comparable parts of $\nu_{\rm mol}$ and the radiative viscosity
\bea
\nonumber
\nu = \nu_{\rm mol} + \nu_{\rm rad},\ \ \mbox{where}\ \ \nu_{\rm rad} = \frac{4aT^4}{15c\varkappa\rho^2}.
\eea
For the reader's convenience, Table~\ref{tab:tab1} shows the correspondence between similar parameters 
in the oceanic and RGB cases as well as their characteristic values.
For the oceanic case, we have used the data compiled by \cite{k87} (also, see Table~1 in the review by \citealt{k03}). For the RGB case, the data are taken from
our bump luminosity model that has a mass $M=0.83\,M_\odot$ and heavy-element mass fraction $Z=0.0005$.
They refer to a point in the $^3$He-burning region of the HBS shell at which $\nabla_\mu$ is negative and has its maximum absolute value
(for model details, see Section~3.3). 

From the lower panel of Fig.~\ref{fig:f1}, it is seen that $Kk^2\gg 2\sigma$ and $Kk^2\gg 2\nu k^2\sim\nu_{\rm mol}k^2$ in the RGB case. 
This allows one to approximate the cubic equation (\ref{eq:dispers}) by the quadratic
\bea\nonumber
2Kk^2\sigma^2 + \left[K(2\nu+\nu_{\rm mol})k^4+2g\beta\frac{\partial\mu_0}{\partial z}(R_\rho-1)\right]\sigma \\
- g\beta\frac{\partial\mu_0}{\partial z}(K-R_\rho\nu_{\rm mol})k^2+\nu K\nu_{\rm mol}k^6 = 0.
\label{eq:quadr}
\eea
For short horizontal wave numbers, such that
\bea\nonumber
k^4\ll g\beta\frac{\partial\mu_0}{\partial z}\frac{(K-R_\rho\nu_{\rm mol})}{\nu K\nu_{\rm mol}},\ \ \mbox{and}\ \
k^4\ll 2g\beta\frac{\partial\mu_0}{\partial z}\frac{(R_\rho - 1)}{K(2\nu +\nu_{\rm mol})},
\eea
a positive solution of (\ref{eq:quadr}) is simplified to
\bea\nonumber
\sigma\approx\frac{1}{2}\,\frac{(K-R_\rho\nu_{\rm mol})}{(R_\rho -1)}\,k^2 =
2\pi^2\,\frac{(K-R_\rho\nu_{\rm mol})}{(R_\rho -1)}\,\frac{1}{d^2}.
\eea
The last expression describes very well the decline of the finger's growth rate with an increase of its diameter at $d\gg d_{\rm max}=d(\sigma_{\rm max})$.
It can be used to estimate a thermohaline diffusion coefficient $D_\mu\approx wl\approx\frac{1}{2}\sigma l^2$. This leads to
the well-known results
\bea
D_\mu\approx \frac{1}{4}\,\frac{(K-R_\rho\nu_{\rm mol})}{R_\rho -1}\,k^2l^2 = 
C\frac{\nabla_\mu}{\nabla_{\rm rad}-\nabla_{\rm ad}-\nabla_\mu}\left(1-\frac{\nu_{\rm mol}}{K}\,\frac{\nabla_{\rm rad}-\nabla_{\rm ad}}{\nabla_\mu}\right)\,
a^2K,
\label{eq:dmu}
\eea
where $C = C_{\rm Kunze}=\pi^2$ and
\bea\nonumber
\nabla_{\rm rad} = \left(\frac{\partial\ln T_0}{\partial\ln P}\right) = \frac{3}{16\pi Gac}\,\frac{\varkappa P}{T^4}\,\frac{L_r}{M_r}
\eea
is the radiative temperature gradient, $M_r$ and $L_r$ being the mass and luminosity at the radius $r$ in the star (assuming spherical symmetry).
The first of the expressions (\ref{eq:dmu}) indicates that the double-diffusive instability develops only
when $1 < R_\rho < 1/\tau$, where $\tau = \nu_{\rm mol}/K = k_S/k_T$ is the inverse Lewis number. This result is known since
the pioneering work of \cite{s60}. Its discussion in regard to the RGB thermohaline mixing has been presented by \cite{dp08b}
(also, see \citealt{v04}).
\cite{u72}, who neglected the influence of salt-fingering on the ambient $T$- and $\mu$-gradients, had obtained an expression for
$D_\mu$ with the constant $C = C_{\rm Ulrich} = 8\pi^2/3$. He did not retain the impeding factor in the parentheses either.
It is Ulrich's formula that has been employed by \cite{chz07}. \cite{kea80} used a different physical approach to arrive at a similar
expression for $D_\mu$ with $C = C_{\rm Kipp} = 12$. In addition, they came to the conclusion that, because of their strong interactions
with the surrounding medium, the perturbed blobs of fluid did not have a chance to become salt-fingers, so that their aspect ratio
should be as small as $a = l/d\approx 0.5$.

It is useful to estimate the diameter of the fastest growing fingers $d_{\rm max}$ as a function of other parameters.
This can be done by analyzing the dependence of the positive root of equation (\ref{eq:quadr}) on the wave number $k$.
It turns out that a sufficiently good approximation for both the oceanic and RGB cases is
\bea
d_{\rm max}\approx 2\pi\sqrt[4]{\frac{\nu k_T}{N_T^2}},
\label{eq:dmax}
\eea
where $N_T^2$ is the square of the unperturbed $T$-component of the Brunt-V\"{a}is\"{a}l\"{a} (buoyancy) frequency. For the RGB case,
we have $N_T^2 = (\nabla_{\rm ad}-\nabla_{\rm rad})g/H_P$, $H_P$ being the pressure scale height. We find that $N_T^2\approx 1.5\times 10^{-4}$\,s$^{-2}$ 
at the same location in our RGB model for which the other data are listed in Table~\ref{tab:tab1}. Its substitution into (\ref{eq:dmax}) together with
the values of $\nu$ and $k_T = K$ gives $\log_{10}d_{\rm max}\approx 4.4$ which is very close to the position of
the maximum on the solid curve in the lower panel of Fig.~\ref{fig:f1}. 

Solid curves in Fig.~\ref{fig:f1} demonstrate that the diameter and velocity of the fastest growing salt fingers in
the oceanic ($R_\rho = 1.6$) and RGB cases are 
$d_{\rm max}^{\rm \,Ocean}\approx 3$ cm, $w_{\rm max}^{\rm Ocean}\approx 0.003\, a_{\rm max}^{\rm Ocean}$\,cm\,s$^{-1}$,
and $d_{\rm max}^{\rm RGB}\approx 0.3$ km, $w_{\rm max}^{\rm RGB}\approx 2\, a_{\rm max}^{\rm RGB}$\,cm\,s$^{-1}$, where
$a_{\rm max}^{\rm Ocean}$ and $a_{\rm max}^{\rm RGB}$ are the (unknown) finger aspect ratios for the corresponding cases.
The major uncertainty in the linear theory is the parameter $a$. Its value can be estimated only by numerical simulations that
directly solve the original non-linear equations (\ref{eq:bouss1}\,--\,\ref{eq:bouss3}). The double-diffusive instability is a primary
instability. It is responsible for the initial growth of salt fingers. What will happen later on and, in particular, how far
the fluid blob will travel vertically, forming a salt finger before its trajectory is bent or the finger gets destroyed by 
its interactions with other fingers and surrounding medium, is determined by secondary instabilities. It is this problem that
we are going to address in the next section.

\section{2D Numerical Simulations of Thermohaline Convection}
\label{sec:numeric}

\subsection{Basic Equations and a Method of Their Solution}

In the Boussinesq approximation, the continuity equation is simplified to $(\nabla,\mathbf{v}) = 0$. It can be automatically satisfied
by representing the velocity vector with a stream function $\psi$, such that
\bea
\mathbf{v} = (u,w) = (-\frac{\partial\psi}{\partial z},\,\frac{\partial\psi}{\partial x}),
\label{eq:stream}
\eea
where $u$ is the velocity (horizontal) $x$-component. After the substitution of (\ref{eq:stream}) into the original system (\ref{eq:bouss1}\,--\,\ref{eq:bouss3}),
we obtain a new system of Boussinesq equations that contain only scalar functions and their derivatives
\bea
\label{eq:nonlin1}
\frac{\partial\,\nabla^2\psi}{\partial t} & = & -J(\psi,\nabla^2\psi) + Pr\frac{\partial}{\partial x}(T'-\frac{S'}{R_\rho}) + Pr\nabla^4\psi, \\
\label{eq:nonlin2}
\frac{\partial T'}{\partial t} & = & -J(\psi,T') - \frac{\partial\psi}{\partial x} + \nabla^2T', \\
\label{eq:nonlin3}
\frac{\partial S'}{\partial t} & = & -J(\psi,S') - \frac{\partial\psi}{\partial x} + \tau\nabla^2S',
\eea
where $J(a,b) = \frac{\partial a}{\partial x}\frac{\partial b}{\partial z} - \frac{\partial a}{\partial z}\frac{\partial b}{\partial z}$ is
the Jacobian. Unlike the quantities $\delta T(x,y,t)$ and $\delta S(x,y,t)$ in equations (\ref{eq:lin1}\,--\,\ref{eq:lin3}), the variances
$T'$ and $S'$ describe deviations of temperature and salinity from $T_0(z)$ and $S_0(z)$ not only in the horizontal plane but also
along the radius, so that
\bea\nonumber
T(x,z,t) = T_0(0) + \left(\frac{\partial T_0}{\partial z}\right)\,z + T'(x,z,t), \\ \nonumber
S(x,z,t) = S_0(0) + \left(\frac{\partial S_0}{\partial z}\right)\,z + S'(x,z,t), \\ \nonumber
\eea
where $T_0(0)$ and $S_0(0)$ are constants measuring the temperature and salinity in the initial unperturbed state at an
arbitrarily chosen vertical position that corresponds to $z=0$.
Equations (\ref{eq:nonlin1}\,--\,\ref{eq:nonlin3}) have been non-dimensionalized using
$d$ and $d^2/k_T$ as the length and time units, where $d = d_{\rm max}/2\pi$, and dividing $T$ and $S$ by $(\partial T_0/\partial z)\,d$
and $(\partial S_0/\partial z)\,d$, respectively.

We solve the system of equations (\ref{eq:nonlin1}\,--\,\ref{eq:nonlin3}) numerically for a 2D doubly-periodic Cartesian domain by employing a computer code
kindly provided by Bill Merryfield. According to him, the code uses a Fourier collocation method with dealiasing that follows
the $2/3$ rule (\citealt{cea88}). Integration in time is via the leapfrog method, with the time-splitting instability
damped by applying a Robert filter with the parameter $0.002$ (e.g., \citealt{wea89}). Dissipation terms are represented by
exponential integration factors. The linear portion of the code was tested by comparing the evolution of small initial
perturbations to predictions of linear stability theory (\citealt{sch79}). To check the non-linear portion, the code was tested
for conservation of temperature variance, salinity variance, energy and enstrophy after the dissipation, buoyancy, and
background gradient terms had been removed. Initial $S'$ is specified by selecting Fourier coefficients from
a bi-Gaussian distribution and scaling by $k^{-2}$, where $k$ is the magnitude of wave vector $\mathbf{k}$.
The variance of initial non-dimensional $S'$ is normalized to $0.1$. Initial values of $T'$ and $\psi$ are set to zero.

\subsection{The Oceanic Case}

For test purposes, we have first reproduced one of the results presented by Merryfiled \& Grinder (2000) in their unpublished paper
(personal communication from Bill Merryfield). It corresponds to the oceanic case with the density ratio $R_\rho = 1.6$
(Table~\ref{tab:tab1}). As output, the code directly gives the effective fingering salt and heat diffusivities,
$D_S$ (denoted as $D_\mu$ in the RGB case) and $D_T$, normalized by $k_T$ ($K$). This is achieved by a procedure that first determines the fingering
salt ($\mu$) and heat fluxes by spatially and temporally averaging the products $wS$ and $wT$ and then dividing them by $\partial S_0/\partial z$
($\partial\mu_0/\partial z$) and $\partial T_0/\partial z$, respectively, after the fingers have attained statistical equilibrium.
The horizontal size of the solution domain is initially chosen to be wide enough, $L_x = L_z = 16\,d_{\rm max}$, so that
the averaging does not produce large-amplitude fluctuations. The blue and red solid curves in the lower panel of Fig.~\ref{fig:f2}
show transitions of the ratios $D_S/k_T$ and $D_T/k_T$ to their equilibrium values. The upper panel gives
a snapshot of salt fingers in the statistical equilibrium. This solution has been obtained with a $1024\times 1024$ resolution for
the salinity, while the code always uses a $2\times 2$ times lower resolution for $T$ and $\psi$. The dashed blue line
in the lower panel plots the constant $D_S/k_T = 2\pi^2a^2/R_\rho$, where $a=3$. It approximates the linear thermohaline
diffusion coefficient (\ref{eq:dmu}) for $C = 2C_{\rm Kunze} = 2\pi^2 = 19.7$, given that $\tau\ll 1$ (Table~\ref{tab:tab1}).
Here, as well as in the next section, we will compare our numerical results for $D_S$ and $D_\mu$ with those
predicted by the linear theory (eq. \ref{eq:dmu}) using $C = 2C_{\rm Kunze}$.
With this choice, $C$ lies approximately in the middle between $C_{\rm Kipp} = 12$ and $C_{\rm Ulrich} = 8\pi^2/3 = 26.3$. 
The comparison gives an effective value of the finger aspect ratio that, after being substituted in (\ref{eq:dmu}),
leads to a value of the linear diffusion coefficient of the same order of magnitude as the one derived from
our numerical simulations. In particular, we have obtained $a\approx 3$ for the test case. This estimate, as well as
the salinity patterns in the upper panel of Fig.~\ref{fig:f2}, confirm that we really deal with (i.e., reproduce numerically) vertically elongated
structures (fingers) in the oceanic case.

Note that, although not perfect, the 2D numerical simulations of salt-fingering in the ocean have succeeded in
producing both the salt and heat fluxes compatible with those measured by \cite{slsch99} in the North Atlantic
tracer release experiment (e.g., Fig.~9 in the review by \citealt{k03}).

\subsection{The RGB Case}

We have employed two stellar evolution codes to compute bump luminosity models of low-mass RGB stars: our own code,
the most recent version of which is described by \cite{dea10}, and the MESA code available at {\tt http://mesa.sourceforge.net}.
The combinations of mass, heavy-element and helium  mass fractions for which the models have been computed with our
code are $(M/M_\odot,Z,Y) = (0.83,0.0005,0.24),\ (0.83,0.001,0.24)$, and $(0.83,0.002,0.24)$, while the models produced
with the MESA code have $(M/M_\odot,Z,Y) = (0.85,0.0001,0.24),\ (0.83,0.001,0.24)$, and $(1.5,0.0188,0.27)$.
The MESA lowest-metallicity model reproduces conditions at which the RGB extra-mixing is thought to be
most efficient, according to both observations and theory (e.g., \citealt{msb08}). 
The parameters of the common model, $M=0.83\,M_\odot$ and $Z=0.001$, are close to those of the
field stars with known Hipparcos parallaxes for which evolutionary abundance changes have been found on the upper RGB
by \cite{grea00}. Finally, the solar-metallicity model has a mass and composition typical for the so-called
Li-rich giants that are located close to the bump luminosity (\citealt{chb00}) and in which Li is believed to be produced 
in a large amount by enhanced extra-mixing (\citealt{dh04}).

Sharp declines of both the Li abundance and carbon isotopic ratio at the bump luminosity in the majority of low-mass RGB stars (e.g.,
\citealt{grea00,sh03,dom10}) demonstrate that the RGB extra-mixing starts to operate and becomes very efficient already
at this evolutionary phase. Hence, if it is actually caused by the $^3$He burning and its associated thermohaline convection,
the latter should already be present in the bump luminosity models in which the HBS has erased the chemical composition discontinuity
left behind by the base of convective envelope at the end of the first dredge-up and, as a result, the $\mu$ depression
produced by the $^3$He burning is now a prominent feature on the otherwise uniform $\mu$ profile outside the HBS.

MESA is a state-of-the-art stellar evolution code that even allows non-stop computations through the helium core flash
toward the end of the asymptotic giant branch (AGB) evolution for low-mass stars. We will use the MESA models as
a background in our post-processing 1D simulations of the RGB extra-mixing. A comparison
shows that the common models in the two sets of computations have very similar structures. 
Curves in the lower panel of Fig.~\ref{fig:f3} show the density ratio profiles
in the  vicinity of the HBS in our models. Their minima are close to the value of $R_\rho = 1700$ that
is used in our 2D numerical simulations of the RGB thermohaline convection. They correspond to the maximum negative values of $\nabla_\mu\approx -10^{-4}$ that are
reached a short distance outside of the $\mu$ depression floor, where $\nabla_\mu = 0$ and $R_\rho = \infty$.
The approximate values of other relevant parameters at the location of the minimum $R_\rho$ are listed in Table~\ref{tab:tab1}.
They do not vary appreciably between the models with different values of $Z$.

\begin{figure}
\epsffile [60 150 580 595] {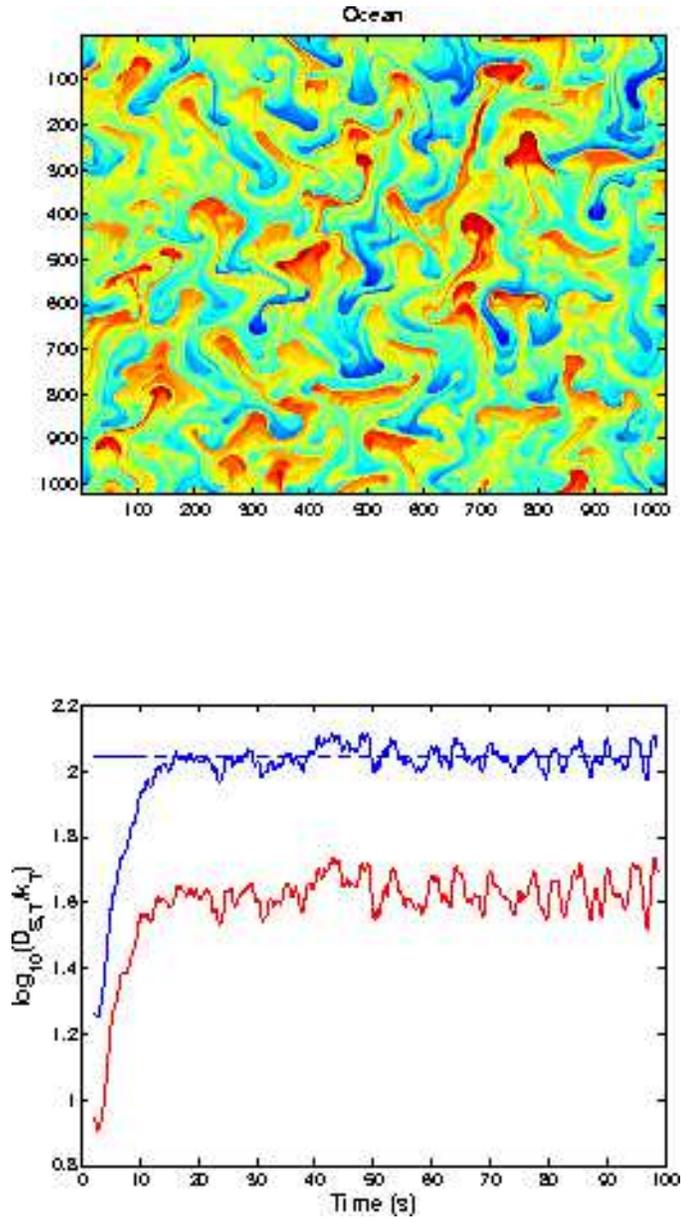}
\caption{A snapshot of our $1024\times 1024$ numerical simulations of the oceanic salt-fingering at the equilibrium state (upper panel). A change of color from blue to red
         corresponds to an increase of salinity. Blue and red curves in the lower panel show the diffusion coefficients for the salinity and heat
         salt-fingering transport in units of the microscopic thermal diffusivity. The blue dashed line approximates the equilibrium value
         of $D_S/k_T$.
         }
\label{fig:f2}
\end{figure}

\begin{figure}
\epsffile [60 260 580 745] {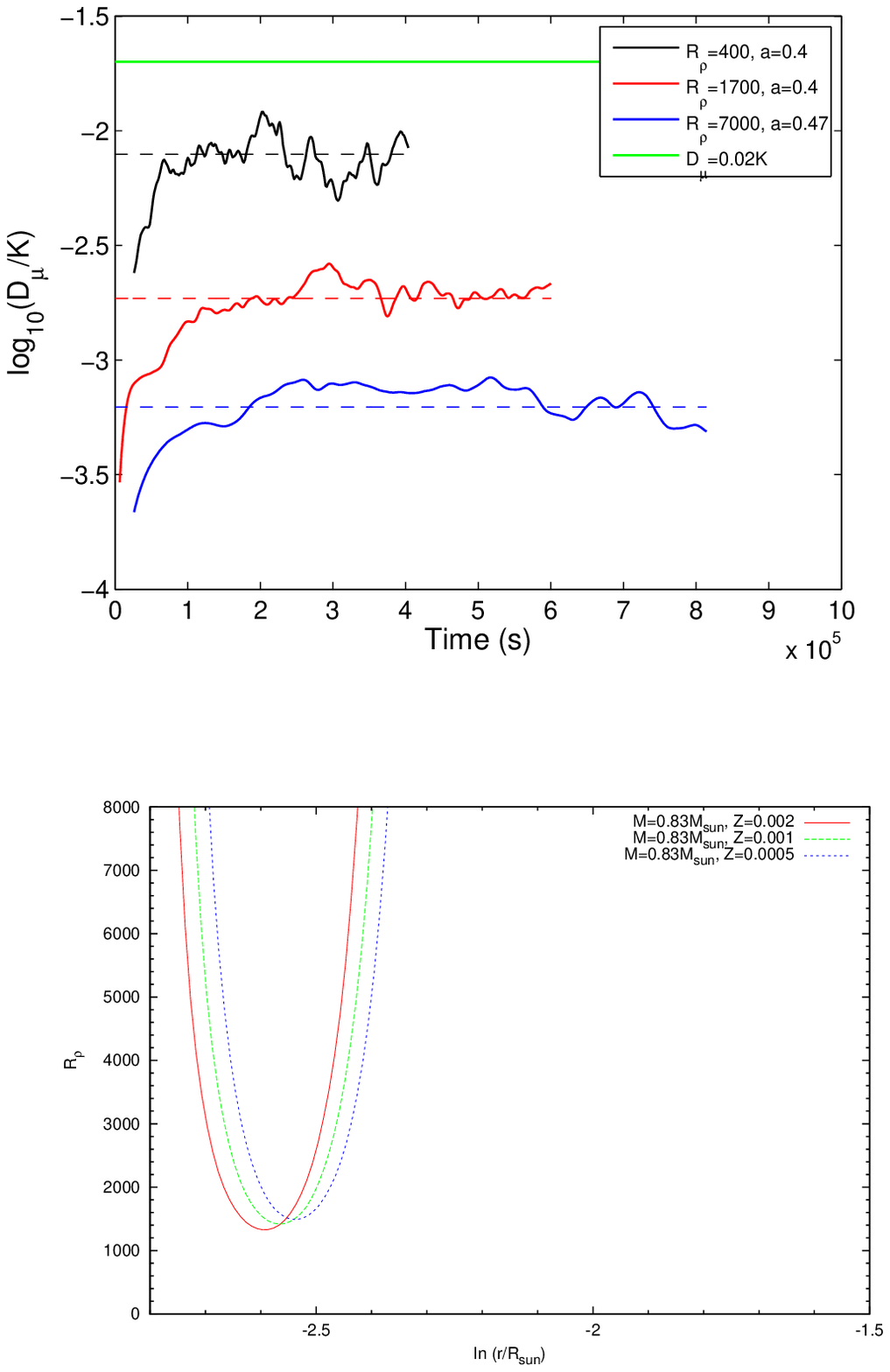}
\caption{Upper panel shows thermohaline diffusion coefficients (oscillating curves) and their corresponding equilibrium
         values (dashed lines) obtained in our 2D numerical simulations for the density ratios
         $R_\rho = 400$, 1700, and 7000, which are close to those found in the radiative zones of
         RGB stars (lower panel). The solid green line in the upper panel gives the empirically constrained rate of the RGB extra-mixing.
         The diffusion coefficients are divided by the thermal diffusivity. In the upper panel, the inset also
         shows the effective finger aspect ratios ($a$) that produce the corresponding equilibrium values (dashed lines)
         when $C=2C_{\rm Kunze} = 2\pi^2$ is substituted in the expression (\ref{eq:dmu}).
         }
\label{fig:f3}
\end{figure}

\cite{dp08b} have shown that evolutionary changes (those which vary with luminosity)
of the surface abundances of Li, C, N, and of the $^{12}$C/$^{13}$C ratio above the bump luminosity in the
field RGB stars with $Z\approx 0.002$ from the study of \cite{grea00} can be reproduced theoretically
if extra-mixing in their radiative zones is modeled using a diffusion coefficient
$D_{\rm mix}\approx 0.02K$. This gives an order-of-magnitude estimate of the rate of extra-mixing that
has to be provided by its correct physical mechanism. The value of $D_\mu/K = D_{\rm mix}/K = 0.02$ is plotted in Fig.~\ref{fig:f3}
(green line). It is to be compared with the red curve in the same figure that has been computed using the same method and resolution
which were used to prepare the blue curve in Fig.~\ref{fig:f2} for the oceanic case but now for the set of parameters corresponding to the RGB case
(Table~\ref{tab:tab1}). To find out how the numerical value of $D_\mu$ depends on the density ratio, we have repeated
the computations for $R_\rho = 400$ (black curve) and $R_\rho = 7000$ (blue curve). The results presented in Fig.~\ref{fig:f3}
can be summarized as follows. By fitting the three curves with their corresponding linear thermohaline diffusion coefficients
(eq. \ref{eq:dmu}) (dashed lines of the same colors), we determine very similar values of the effective finger aspect ratio, $a\approx 0.4$\,--\,$0.5$.
This means that the effective salt-fingering $\mu$-diffusivity can indeed be approximated by (\ref{eq:dmu}) or, in other words, that
$D_\mu/K\propto 1/R_\rho\propto |\nabla_\mu|$, at least within the investigated parameter range. 
Second, the value of $a$ resulting from our direct 2D numerical simulations
turns out to be surprisingly close to the value of $a=0.5$ advocated by \cite{kea80}, whereas it is an order of magnitude smaller than
the value assumed by \cite{u72}. Third, the equilibrium value of $D_\mu$ for our standard RGB case 
(i.e., for the values of relevant parameters in the vicinity of the HBS presented in Table~\ref{tab:tab1}) is nearly an order of magnitude smaller
than the value of $D_{\rm mix} = 0.02K$ that satisfies the observational data (compare the red and green lines). Note that the latter was assumed to be constant
during the upper RGB evolution, whereas $D_\mu$ will obviously decline with time along with an evolutionary decrease of
the envelope $^3$He abundance, hence $|\nabla_\mu|$, which will make it even more difficult for the $^3$He-driven thermohaline convection
to comply with observations (see the next section). Finally, the effective thermohaline heat diffusivity is found to be
negligibly small compared to the radiative thermal diffusivity $K$ in all the cases. Consequently, it will not
influence the thermal structure of the radiative zone.

\begin{figure}
\epsffile [60 150 580 595] {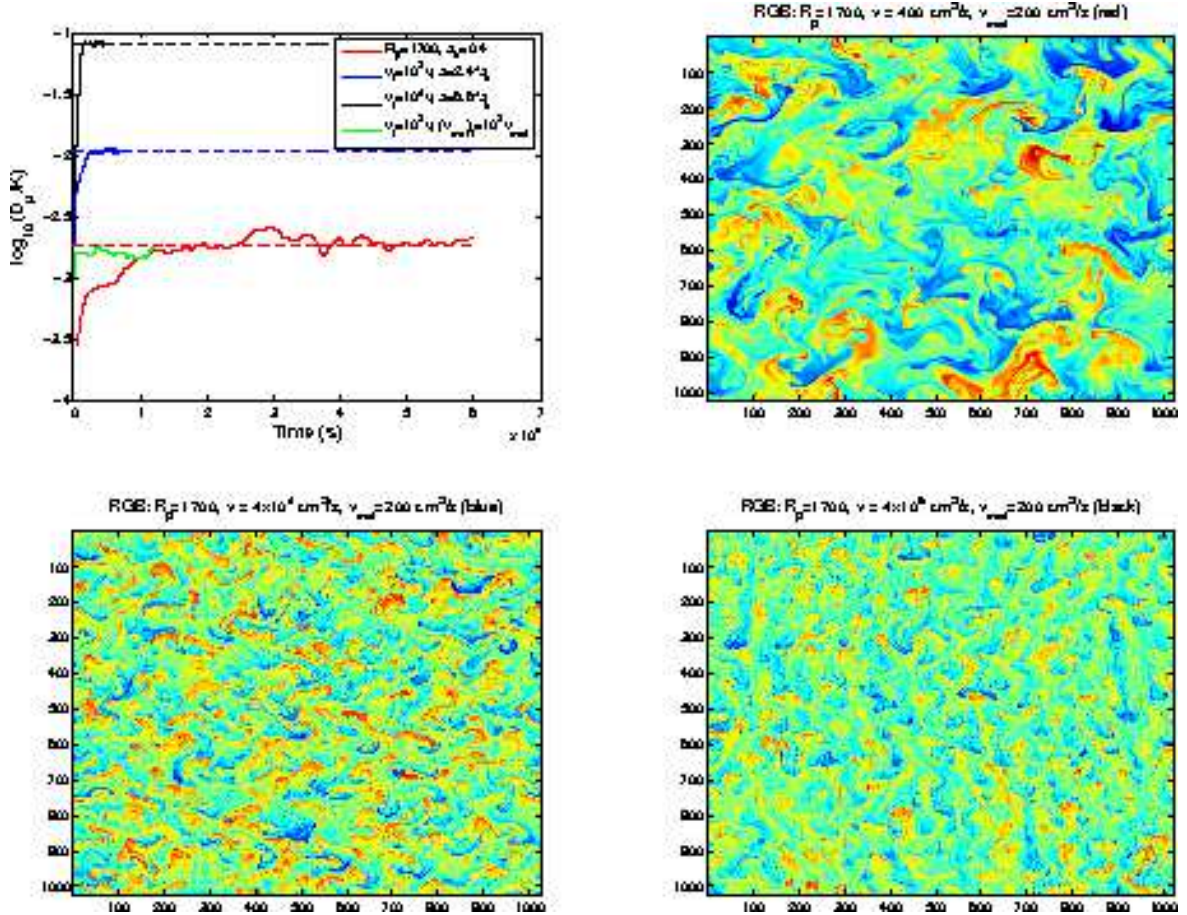}
\caption{Snapshots and results of our 2D numerical simulations of thermohaline convection for the RGB case with the same density ratio
         $R_\rho = 1700$ but for different values of viscosity and mixing rate. Upper-right panel: the standard RGB case
         (see Table~\ref{tab:tab1}). Lower-left and lower-right panels: the viscosity has been increased by the factors of
         $10^2$ and $10^4$ (assuming it is now of turbulent origin). Upper-left panel: the diffusion coefficients and effective
         finger aspect ratios corresponding to the cases presented in the other three panels. The green curve has the same
         parameter set as the blue curve except that its corresponding mixing rate (molecular diffusivity $\nu_{\rm mol}$)
         has also been increased by the factor of $10^2$.
         }
\label{fig:f4}
\end{figure}

The red curve from Fig.~\ref{fig:f3} is also plotted in the upper-left panel of Fig.~\ref{fig:f4}, while the upper-right panel
shows a snapshot of its corresponding salinity ($\mu$ in this case) field at equilibrium. A comparison of the latter panel with the upper panel of
Fig.~\ref{fig:f2} leads us to the conclusion that, unlike the oceanic case, there are no visible vertically elongated
structures (fingers) in the RGB case. This is the reason why we get $a < 1$ in the second case. The $\mu$-field patterns in
the upper-right panel of Fig.~\ref{fig:f4} instead resemble disorganized turbulent-like convection. A possible explanation of
the difference in the salinity-field structures between the oceanic and RGB case can be given on a basis of results of the analysis of
the equilibration of the salt-finger instability reported by \cite{r10}. First of all, it should be noted that the difference between
the two cases is evidently caused by their corresponding different values of the three dimensionless parameters in equations
(\ref{eq:nonlin1}\,--\,\ref{eq:nonlin3}): $R_\rho$, $Pr$, and $\tau$ (the last three rows in Table~\ref{tab:tab1}). 
\cite{r10} replaced the original non-linear equations
by their weakly non-linear analogs using asymptotic expansions in which $\varepsilon = 1/R_\rho - \tau\ll 1$ and then he solved them
numerically for a wide range of $Pr$ employing a computer code similar to that used by us. As a result,
he has isolated two dominating mechanisms of salt-fingering equilibration. These are the triad interactions between
various growing modes (e.g., \citealt{v06}) and the adverse action of vertical shear, spontaneously developing as a result of secondary salt-finger
instability. \cite{r10} has come to the conclusion that, although both processes are essential, it is the shear instability that
becomes critical at $Pr\ll 1$. Under these circumstances, the growth of fingering salinity and heat fluxes is limited by
the energy dissipation sufficient to balance the buoyancy forcing generated by the double-diffusive (primary) instability
(\citealt{sh95}). The fluxes generally decrease with the Prandtl number, although their dependence on $Pr$
is non-monotonic. 

To verify these predictions for the RGB case, we have repeated our computations with the viscosity ($Pr\propto\nu$)
artificially increased by factors of $10^2$ and $10^4$. The effective $\mu$-diffusivities 
for these test cases, as well as their approximations by the linear
diffusion coefficient (\ref{eq:dmu}), are plotted with blue and black solid and dashed curves
in the upper-left panel of Fig.~\ref{fig:f4}. We see that the increase of $Pr$ indeed results in an increase
of $D_\mu$, the latter being proportional to the $\mu$-flux. The linear approximation of $D_\mu$ for $\nu_{\rm t} = 10^4\,\nu=4\times 10^6$ cm$^2$\,s$^{-1}$
yields the effective finger aspect ratio $a\approx 6.6\,a_0=2.6$,
where $a_0=0.4$ for the standard RGB case. We use a subscript ``t'' in the expression for viscosity to emphasize the fact that this cannot
be a microscopic viscosity anymore. Instead, it could be a turbulent viscosity like the one associated with rotation-driven
instabilities. When $\nu = 4\times 10^6$ cm$^2$\,s$^{-1}$, the $\mu$-field patterns, like the value of $a$, 
appear to almost be identical to those obtained for the oceanic case 
(compare the lower-right panel of Fig.~\ref{fig:f4} with upper panel of Fig.~\ref{fig:f2}).
The case of $\nu_{\rm t} = 10^2\,\nu = 4\times 10^4$ cm$^2$\,s$^{-1}$ corresponds to a transition from the $\mu$-field structure 
at $\nu = 4\times 10^6$ cm$^2$\,s$^{-1}$ that looks more or less
organized in the vertical direction to the structure resembling a turbulent-like convection at $\nu = 400$ cm$^2$\,s$^{-1}$. The most interesting pattern
seen in the lower-left panel of Fig.~\ref{fig:f4} is something like a vertically varying sinusoidal 
shear. This is most likely to be a manifestation of the mean field effects (a secondary instability) that become critical at low $Pr$,
according to \cite{r10}. It is also interesting to compare our snapshot panels corresponding to the decreasing $\nu$ with 
the sequence of $T$-field snapshots presented in Fig.~2 by \cite{sh95}. They look very similar. However, the important difference is that
Shen shows a time sequence along which the secondary instability, the one that leads to the equilibrium salt-fingering convection,
is developing, whereas our sequence shows the equilibrium states already achieved for different values of $Pr$, at the highest of which
the secondary instability appears to be strongly suppressed.

Our results in the upper-left panel of Fig.~\ref{fig:f4}, in particular, the one represented by the black curve, 
could be used to support the hypothesis that
the $^3$He-driven thermohaline convection is the main RGB extra-mixing mechanism, provided that
a sufficiently strong turbulent viscosity, say of rotational origin (e.g., \citealt{pea06}), would be present in the radiative zones of RGB stars.
However, such a speculation has to admit (and explain how it is possible) that the turbulence enhances only the viscosity but not
the $\mu$-diffusivity. Indeed, when we increase both $\nu$ and $\nu_{\rm mol}$ by the same factor,
the effective thermohaline $\mu$-diffusivity returns close to its standard value (the green curve in the upper-left
panel of Fig.~\ref{fig:f4}). This is not surprising because an increase of $\nu_{\rm mol}$ ($k_S$ in the oceanic case)
reduces the buoyancy force by making it faster for the difference in the chemical composition between rising and sinking blobs
to be smoothed out horizontally (see the second term in the parentheses in eq. \ref{eq:dmu}). 
Furthermore, if we accept the hypothesis that turbulence in stellar radiative zones should be highly anisotropic,
with its associated horizontal viscosity components strongly dominating over that in the vertical direction (\citealt{z92}),
then the above speculation becomes even less likely, again provided that the chemical composition transport is
accelerated by the turbulence proportionally to that of momentum.

\section{Post-Processing 1D Simulations of the RGB Thermohaline Mixing}
\label{sec:1d}

The results of our 2D numerical simulations of thermohaline convection in the vicinity of the HBS in the bump luminosity
RGB model predict the effective salt-finger aspect ratio $a=0.4$\,--\,$0.5$ that is an order of magnitude smaller than
the one advocated by \cite{u72}, the latter being needed to reproduce the \cite{grea00} observational data, according to \cite{chz07}.
The small value of $a$ in the RGB case compared to its much higher value in the oceanic case is due to
the more favourable conditions for the development of secondary instabilities, that strongly limit the growth of salt fingers,
at low Prandtl numbers. 

However, our simulations are far from being perfect. Indeed, first of all, they are
two-dimensional, in which case the oceanic salt-fingering fluxes were found to be underestimated by a factor of two to three compared 
to 3D numerical simulations (\citealt{rs99}). For low Prandtl numbers corresponding to the RGB case, differences between the 2D and 3D
solutions may be even more substantial (\citealt{r10}). Second, our simulations are restricted to a small space domain surrounding the point of minimum $R_\rho$.
Third, they do not take into account either nuclear reactions or a modification of $R_\rho$ by thermohaline mixing, both of which may influence
the growth of salt fingers. Finally, there still remains the possibility that the secondary instability associated with the mean field
effects (vertical shear) can be suppressed in the RGB case, for example, by rotation-induced turbulence, such that the turbulent viscosity
considerably exceeds the rate of turbulent mixing for some reason, which will result in a higher $a$.
Therefore, we have decided to supplement our 2D numerical simulations of the RGB thermohaline convection with
1D computations of the evolutionary changes of the surface chemical composition of RGB stars above the bump luminosity
in which we model the $^3$He-driven thermohaline convection using the linear-theory diffusion coefficient (\ref{eq:dmu}) as in the work of,
e.g. \cite{chz07} and \cite{sea09}. However, given that the diffusion coefficient (\ref{eq:dmu}) is proportional to
an extremely small quantity $|\nabla_\mu|\la 10^{-4}$, which is affected by the mixing itself,
we decided that it would be very sensible to re-mesh the computational grid of full stellar evolution computations,
and to perform our 1D simulations on a fixed mesh in a post-processing way.
Our goal is to see if we can adjust a value of $a$ with which the diffusion coefficient (\ref{eq:dmu})
will really be able to reproduce the $^{12}$C/$^{13}$C, C, and N data of \cite{grea00}. We want to do this also because
there were some discrepancies between the results reported by \cite{chz07} and \cite{sea09}, on the one side, and those
obtained and anticipated by \cite{dp08b}, on the other.

\subsection{Basic Equations and a Method of Their Solution}

\cite{dp08a} have noticed that, in the absence of extra-mixing, the $\mu$-profile in the radiative zone of an RGB star above
the bump luminosity does not change very much if the mean molecular weight is plotted as a function of radius. 
The electron-degenerate He core in the center of the RGB star can be considered as a low-mass white dwarf, whose
radius is known to weakly depend on its mass ($R\propto M^{1/3}$). The He core mass increases with time as
the star climbs the RGB, thanks to the transformation of H into He taking place in the HBS atop the He core.
This needs some fresh H to constantly be conveyed to the HBS from the base of convective envelope through
the radiative zone. Note that, whereas the mass of the radiative zone is very small ($\sim 0.01$\,--\,$0.001\,M_\odot$),
its radius extends from a value of the order of the Earth's radius at the He core boundary to a value of the order of the Sun's radius
at the base of the convective envelope.
A typical radial velocity of the mass inflow between the envelope and HBS, that feeds H to the HBS,
is $\dot{r}\approx -(10^{-4}$\,--\,$10^{-5})$ cm\,s$^{-1}$. Given these facts, it is natural to write and solve the nuclear kinetics equations
for the radiative zone in the Eulerian coordinates
\bea
\frac{\partial y_i}{\partial t} + \dot{r}\,\frac{\partial y_i}{\partial r} =
\left(\frac{\partial y_i}{\partial t}\right)_{\rm nucl} + \frac{\partial}{\partial r}\left(D_\mu\,\frac{\partial y_i}{\partial r}\right),
\label{eq:kin}
\eea
where $y_i$ is the mole-per-gram abundance of the $i^{\rm th}$ nuclide, $D_\mu$ is the thermohaline diffusion coefficient
given by expression (\ref{eq:dmu}) with $C=2C_{\rm Kunze} = 2\pi^2$, and the first term on the right-hand side describes local changes 
of $y_i$ produced by nuclear reactions. 

Distributions of $\rho(r)$, $T(r)$, and other stellar structure parameters in the radiative zone necessary for calculations
of $\dot{r}$, $D_\mu$, and the nuclear term in (\ref{eq:kin}) are taken from a couple of our MESA models separated by
a sufficiently large distance in $\log L/L_\odot$ on the RGB, the first model being located immediately after the bump luminosity,
where the RGB extra-mixing is supposed to commence.
The models are used to interpolate the structure parameters in both $r$ and $t$.
The mass inflow rate can be estimated from the mass and energy conservation relations
\bea
\frac{dr}{dt} = -\frac{1}{4\pi r^2\rho}\,\frac{dM_r}{dt},\ \ \ \ \frac{dM_r}{dt} = \frac{L}{\varepsilon_{\rm CN} X_{\rm e}},
\eea
where $X_{\rm e}$ is the envelope hydrogen mass fraction, $\varepsilon_{\rm CN}\approx 6\times 10^{18}$ erg\,g$^{-1}$ is the energy released
per one gram of hydrogen burnt in the CN-branch of nuclear reactions, and $L$ is the star's luminosity that is assumed to be
constant in the radiative zone above the HBS. A comparison of these relations with real $\dot{r}$-profiles from our RGB models has led to the following corrected
relation 
\bea
\dot{r}\approx -0.6\,\left(\frac{L_\odot}{4\pi R_\odot^2\varepsilon_{\rm CN}}\right)\,\frac{L/L_\odot}{(r/R_\odot)^2\rho X_{\rm e}}
\label{eqrdot}
\eea
that will be used in our 1D simulations.

Our test nucleosynthesis computations have shown that both the location and form of the $\mu$ depression are entirely determined by the reactions of $^3$He
burning and those of the CN-branch of the CNO-cycle. Therefore, in our post-processing RGB extra-mixing simulations, we have included only the following
reactions: $^3$He($^3$He,2p)$^4$He, $^3$He($\alpha,\gamma)^7$Be(p,$\gamma)^8$B($\beta^+\nu_{\rm e}$) 2\,$^4$He, $^{12}$C(p,$\gamma)^{13}$N($\beta^+\nu_{\rm e})^{13}$C,
$^{13}$C(p,$\gamma)^{14}$N, $^{14}$N(p,$\gamma)^{15}$O($\beta^+\nu_{\rm e})^{15}$N, and $^{15}$N(p,$\alpha)^{12}$C. Their rates
have been taken from the NACRE compilation by \cite{aea99}. The abundances of H, $^3$He, $^4$He, $^{12}$C, $^{13}$C, $^{14}$N, and $^{15}$N
from the convective envelope of the MESA bump luminosity model are used as initial uniform conditions in the radiative zone.
Simple boundary conditions, $(\partial y_i/\partial r) = 0$, are applied at the bottom of the HBS, where $X = 10^{-4}$, and at
the star's surface. Mixing in the convective envelope is modeled with the diffusion coefficient $D_{\rm conv} = 10^{12}$ cm$^2$\,s$^{-1}$,
which keeps the envelope composition uniform. The system of equations (\ref{eq:kin}) has been solved numerically by a finite element method using
the COMSOL Multiphysics software package. Initially, we substitute $D_\mu = \nu_{\rm mol}$ into equations (\ref{eq:kin}) and solve them using
the stellar structure parameters only from the first model for a long enough time to determine stationary abundance and $\mu$ profiles
in the background of the radiative zone mass-inflow
(e.g., panels a and c in Fig.~\ref{fig:f5}). After this, we switch on the $^3$He-driven thermohaline mixing
and let it operate as our model evolves by adding the expression (\ref{eq:dmu}) to $D_\mu$ and interpolating
the relevant stellar structure parameters in time and radius between our two RGB models.
Note that the addition of (\ref{eq:dmu}) to the diffusion coefficient introduces an extra non-linearity in the PDEs (\ref{eq:kin}) because 
of the dependence
\bea
D_\mu\propto \frac{\partial\ln\mu}{\partial\ln r} = -\mu\,\sum_i\,(1+Z_i)\,y_i\,\frac{\partial\ln y_i}{\partial\ln r},
\eea
where $Z_i$ is the charge of the $i^{\rm th}$ nuclide. It is this non-linearity associated with extremely small values of the term $(\partial\ln\mu/\partial\ln r)$
plus the frequent re-meshing of the radiative zone needed because the HBS constantly advances in mass that make the implementation of
the $^3$He-driven thermohaline mixing difficult in full stellar evolution computations. The researches who have done such calculations (e.g., \citealt{chz07},
and \citealt{sea09}) did not provide details of their implementations making it impossible to try to reproduce or assess their
results. To eliminate a potential problem associated with
the re-meshing, we solve the PDEs on a fixed mesh. To treat the non-linearity as precisely as possible, we use a large number of mesh points
(approximately 1000) and set up very small tolerances in the time stepping algorithm of the COMSOL code.

\subsection{Solutions for the Low-Metallicity RGB Models}

\begin{figure}
\epsffile [60 180 580 595] {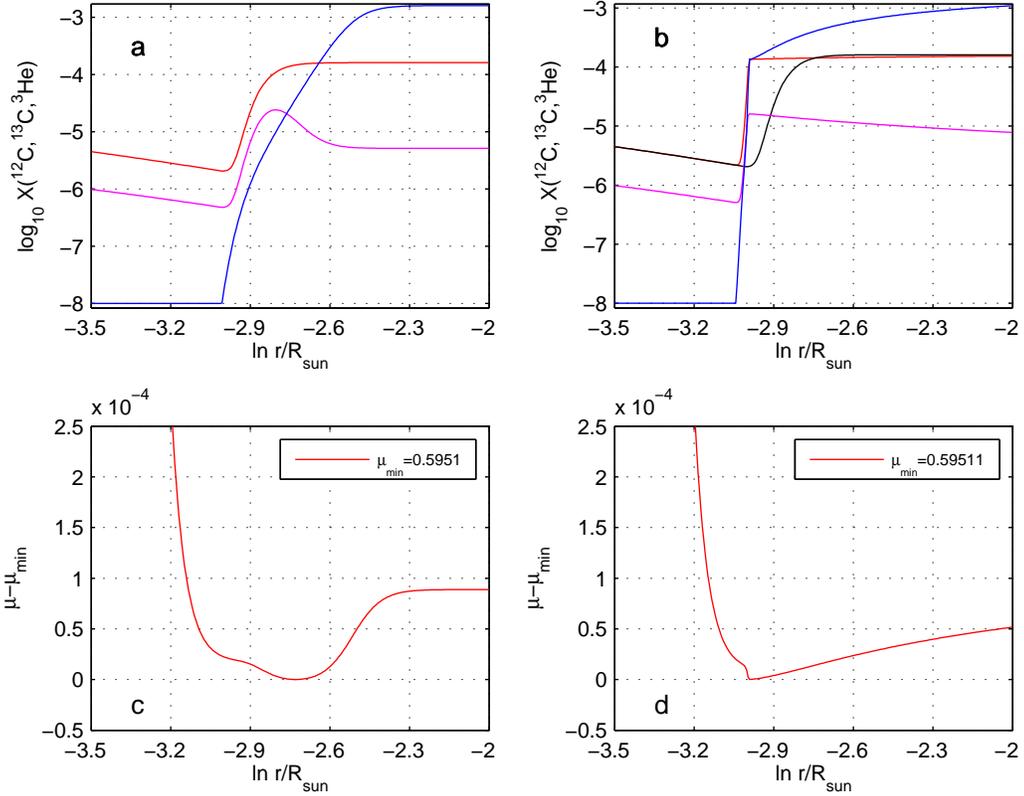}
\caption{The $^3$He, $^{12}$C, and $^{13}$C abundance and $\mu$ profiles in the vicinity of the HBS in the bump luminosity model
         with $Z=0.001$ and $M=0.83\,M_\odot$ computed in the absence of extra-mixing (panels a and c). Panels b and d: the same profiles
         after they have been modified by RGB extra-mixing modeled by the diffusion coefficient (\ref{eq:dmu}) with $a=5$. The black curve in panel b
         is the same as the red curve from panel a.
         }
\label{fig:f5}
\end{figure}

A mean metallicity of upper RGB stars from the \cite{grea00} sample with [C/Fe] abundances clearly demonstrating an evolutionary
decline corresponds to the heavy-element mass fraction $Z\approx 0.0005$, with a dispersion comparable to the mean value.
Their masses as estimated using the Hipparcos parallaxes are centered around $M\approx 0.85\,M_\odot$.
Therefore, we will model the RGB extra-mixing in these stars using MESA RGB models that have $Z=0.001$ and $Z=0.0001$, and the masses $M=0.83\,M_\odot$
and $M=0.85\,M_\odot$, respectively.
Panels a and c in Fig.~\ref{fig:f5} show the $^3$He, $^{12}$C, $^{13}$C, and $\mu$ profiles in the vicinity of
the HBS in the bump luminosity model with $Z=0.001$ in the absence of extra-mixing.
A rule of thumb for estimating a correct depth of the $\mu$ depression is the relation $\Delta\mu\approx\mu^2\Delta X(^3\mbox{He})/6$,
where $\Delta X(^3\mbox{He})$ can be replaced with the envelope $^3$He mass fraction with no harm done.
Applying this rule for $\log_{10}X_{\rm e}(^3\mbox{He})\approx -2.8$ and $\mu\approx\mu_{\rm min}=0.595$ (panels a and c), we obtain an estimate of
$\Delta\mu\approx -9\times 10^{-5}$ that is close to the maximum depth of the $\mu$ depression in panel c.

The minimum of $\mu$ is reached at $\ln\,r_{\rm min}/R_\odot\approx -2.73$, or $r_{\rm min}/R_\odot\approx 0.065$, 
where $r_{\rm min} = r(\mu_{\rm min})$. This is very close to
the values reported in our previous publications (e.g., \citealt{dp08b}). If we draw a vertical line through the point of
$\mu_{\rm min}$ in panel c toward the panel a then we will come to our former conclusion that
the $^3$He-driven thermohaline convection cannot explain the observed decline of [C/Fe] unless it penetrates below 
$\mu_{\rm min}$. It turns out that such overshooting is indeed possible, provided that the thermohaline mixing is
much more efficient than the one predicted by our 2D numerical simulations. Panels b and d in Fig.~\ref{fig:f5}
show the abundance and $\mu$ profiles that are obtained for a case of extra-mixing in which $D_\mu$ has been modeled by equation (\ref{eq:dmu}) 
with $a=5$. Note that this finger aspect ratio is nearly ten times as large as the effective one estimated from Fig.~\ref{fig:f3},
the resulting diffusion coefficient being almost two orders of magnitude higher. Panels b and d correspond to a short time after
the mixing has been switched on. We see that the mixing extends down to the radius $r_{\rm mix}\approx 0.05\,R_\odot$.
Again, this is close to the mixing depth used by \cite{dp08a}. Although the overshooting from $r_{\rm min}$ to
$r_{\rm mix}$ has been produced numerically, it has some physical justification, and can therefore be real.
When the mixing is switched on, it will first steepen the $\mu$-profile immediately to the right from $r_{\rm min}$ (panel c).
Very soon, this will result in the formation of a discontinuity in the $\mu$-profile at $r_{\rm min}$, where material with
a higher $\mu$ overlies material with a lower $\mu$. Such a stratification is also subject to the double-diffusive instability,
like the one with uniform salinity and temperature gradients considered previously. Hence, the material from the right will start
to mix with the material from the left, thus pushing the $\mu$-profile discontinuity to the left and, simultaneously, lowering it.
This process will continue until the point $r_{\rm mix}$ is reached at which the local reduction of $\mu$ by $^3$He burning, with
$^3$He being conveyed by mixing from the envelope, is balanced by the increase of $\mu$ caused by the transformation of H into He
in the CN-cycle.

Fig.~\ref{fig:f6} compares the abundance and $\mu$ profiles
for $a=5$ (panels a and d), $a=6$ (panels b and e), and $a=7$ (panels c and f) in the same bump luminosity model.
Black curves in the three upper panels show the $^{12}$C profile in the model without mixing (the red curve in Fig.~\ref{fig:f5}a).
One can see that the mixing with the higher finger aspect ratio penetrates deeper into the HBS and, at the same time, its diffusion
coefficient is increased proportionally to $a^2$ (eq. \ref{eq:dmu}). However, this does not turn out to strongly change the evolutionary
variations of the surface composition produced by the mixing with the large values of $a$. They are plotted with red ($a=5$),
green ($a=6$), and blue ($a=7$) curves in Fig.~\ref{fig:f7}, where crosses and triangles represent the \cite{grea00}
data, the triangles showing upper limits to the $^{12}$C/$^{13}$C ratio. For comparison, black curves in Fig.~\ref{fig:f7}
correspond to the case of $a=1$. Note that even this small finger aspect ratio still leads to a nearly four times larger
thermohaline diffusion coefficient than the one predicted by our 2D numerical simulations (the red curve in Fig.~\ref{fig:f3}).

\cite{dv03} have demonstrated that an increase of the RGB extra-mixing rate by the factor of two should result in a strong
enhancement of the evolutionary decline of the carbon abundance (see their Fig.~7). An increase of the mixing depth
should also enhance the [C/Fe] depletion, though in a lesser proportion than that of the diffusion coefficient (their Fig.~8).
In the present case, it turns out that, in spite of the fact that the increase of $a$ from 5 to 7 doubles the diffusion coefficient
(\ref{eq:dmu}) and, at the same time, produces deeper mixing, this does not affect the [C/Fe] evolutionary decline
(compare the red and blue curves in Fig.~\ref{fig:f7}b) very much. This difference is explained by a convergence of the diffusion
coefficients (\ref{eq:dmu}) calculated for different but sufficiently large values of $a$ to the same profile (Fig.~\ref{fig:f8}),
as $^3$He gets depleted in the envelope (Fig.~\ref{fig:f7}d). The faster and deeper mixing destroys $^3$He quicker
than the slower and shallower mixing and, because the diffusion coefficient (\ref{eq:dmu}) is indirectly proportional to
the mass fraction of $^3$He left in the envelope (through the dependence $\Delta\mu\propto\Delta X(^3\mbox{He})$), 
the $D_\mu$-profile corresponding to the larger value of $a$ quickly converges to that calculated for the smaller $a$.
In other words, it can be said that the efficient $^3$He-driven thermohaline mixing becomes self-quenching.
On the contrary, the aforementioned results reported by \cite{dv03} were obtained for constant mixing rates.
In connection to this, it should be noted that \cite{dp08b} have reproduced the \cite{grea00} data with
the diffusion coefficient $D_{\rm mix} = 0.02\,K$ also assuming that it does not change with time. Therefore, given that
the coefficient (\ref{eq:dmu}) rapidly decreases with time as a result of the self-quenching (Fig.~\ref{fig:f8}), its initial values have to be
much larger than $0.02\,K$ (Fig.~\ref{fig:f8}a). Hence, the green line in the upper panel of Fig.~\ref{fig:f3}
with which we have compared our 2D numerical simulations of the thermohaline diffusion coefficient should
be increased by approximately one order of magnitude.

The floor of the $\mu$ depression is found to be almost flat in the case of $Z=0.0001$ and $M=0.85\,M_\odot$ (Fig.~\ref{fig:f9}c).
As a result, for the same value of $a=5$, the mixing in this model penetrates deeper (with respect to the $^{12}$C profile in
the unmixed model, shown with black curves in Figs.~\ref{fig:f5}b and \ref{fig:f9}b) than in
the model with the higher metallicity (compare panels b and d in the two figures). 
As in the previous case, the mixing depth increases with $a$ (Fig.~\ref{fig:f10}).
Given that the bump luminosity is higher for the lower metallicity model, the mixing in it starts with a larger initial $K$, hence with a higher $D_\mu\propto K$,
because $K\propto L$ but, at the same time, the mixing has less time to accomplish its task.
Evolutionary changes of the surface chemical composition produced by the mixing are shown in Fig.~\ref{fig:f11}.
We see that the low [C/Fe] ratios in the most luminous RGB stars from the Gratton et al. sample are almost reproduced
with the diffusion coefficient (\ref{eq:dmu}), provided that the finger aspect ratio can reach a value of $a\ga 7$.
Note, however, that this result is obtained for $Z=0.0001$, whereas some of these stars have metallicities closer to $Z=0.001$,
in which case the agreement between our model predictions and the observational [C/Fe] data is worse (Fig.~\ref{fig:f7}b).
Nevertheless, given the model uncertainties, the approximations used in our 1D simulations of the RGB thermohaline mixing, and
the fact that the metallicities of upper RGB stars from the Gratton et al. sample are distributed between $Z\approx 0.001$ and a value of $Z$
close to 0.0001, we find the general agreement between the linear theory with the high finger aspect ratios and observations to be satisfactory. 
Besides, the model confirms the observational inference that the effect of the RGB extra-mixing increases towards lower metallicities.
Thus, the results of our 1D simulations of $^3$He-driven thermohaline mixing in upper RGB stars go along with those reported by \cite{chz07}.
In particular, even our adjusted finger aspect ratio $a=7$ results in a value of the total non-dimensional coefficient $C_{\rm t} = 2\pi^2a^2 = 967$ that
is very close to the value of $C_{\rm t} = 1000$ used by them.

\subsection{Solutions for the Solar-Metallicity RGB Model}

\begin{figure}
\epsffile [60 180 580 595] {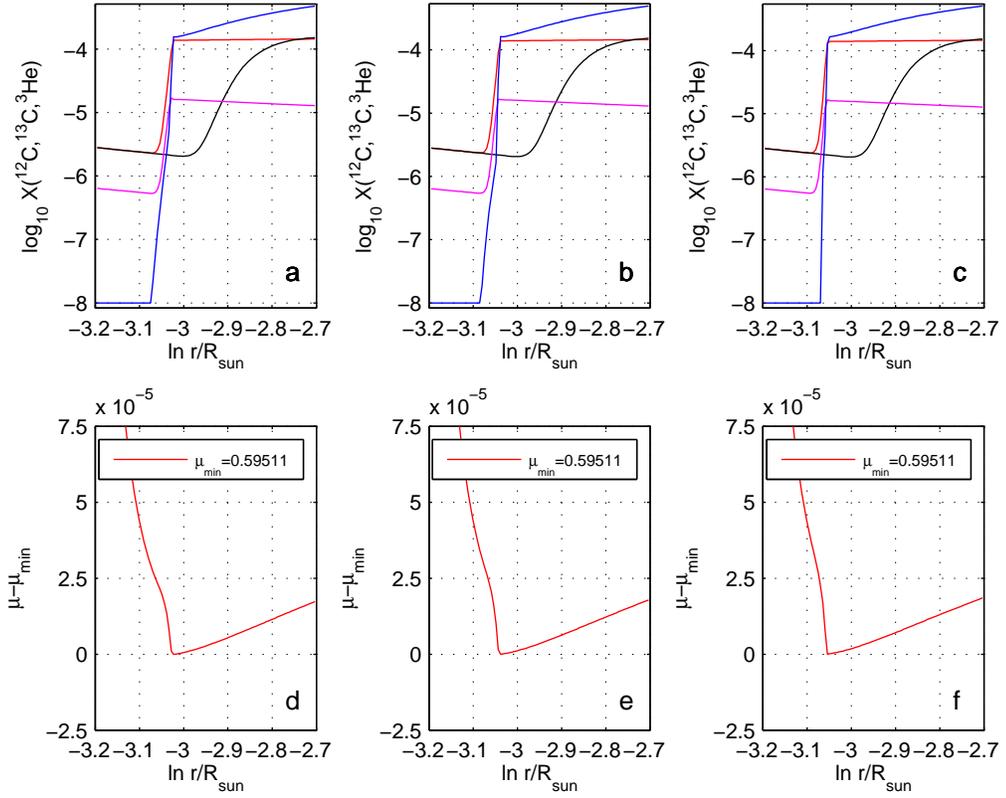}
\caption{The first pair of the upper and lower panels (a and d) is the same as panels b and d in the previous figure, while
         the other two pairs correspond to finger aspect ratios $a=6$ and $a=7$, respectively.
         }
\label{fig:f6}
\end{figure}

\begin{figure}
\epsffile [60 180 580 595] {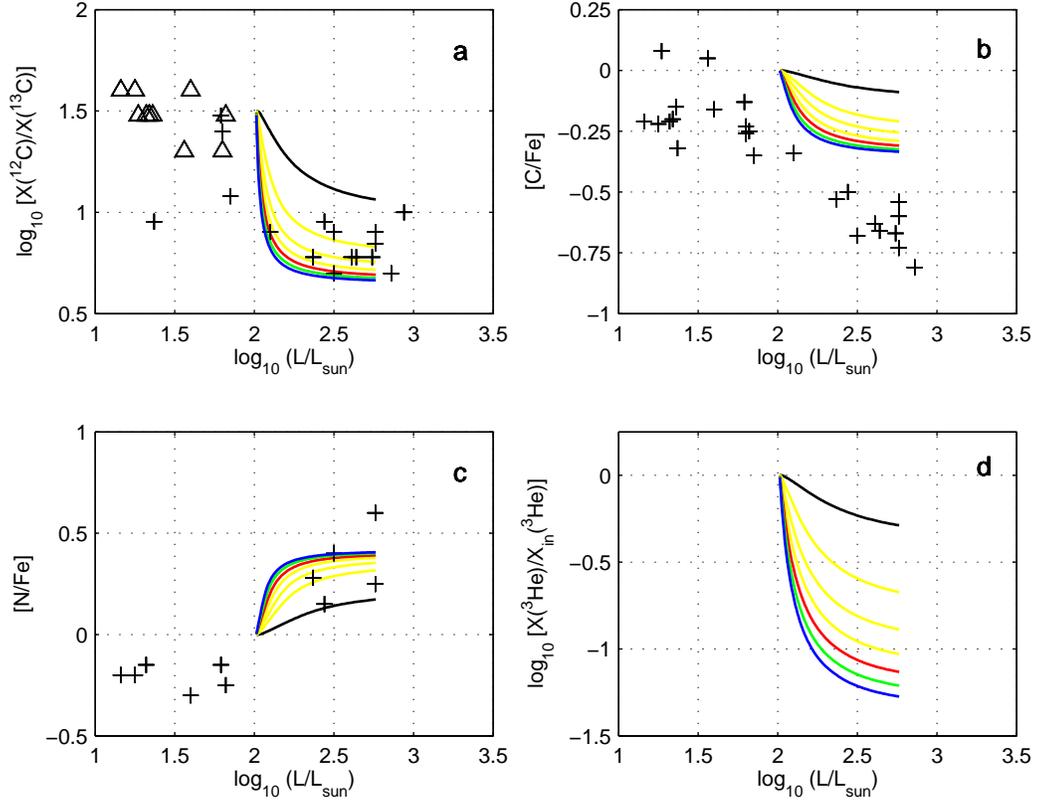}
\caption{The evolutionary changes of the surface $^{12}$C/$^{13}$C ratio (panel a) and the abundances of C (panel b) and N (panel c) in the field low-metallicity
         and low-mass RGB stars studied by \cite{grea00} (crosses and triangles, the latter showing upper limits). Curves represent
         theoretical reproductions by our model of RGB extra-mixing with $Z=0.001$ and $M=0.83\,M_\odot$ in which mixing is described by the diffusion
         coefficient (\ref{eq:dmu}) with $a=1$ (black curves), $a=2$, 3, and 4 (yellow curves), $a=5$ (red curves), $a=6$ (green curves),
         and $a=7$ (blue curves). Panel d shows the unobserved decline of the envelope $^3$He abundances that is predicted theoretically.
         }
\label{fig:f7}
\end{figure}

\begin{figure}
\epsffile [60 180 580 595] {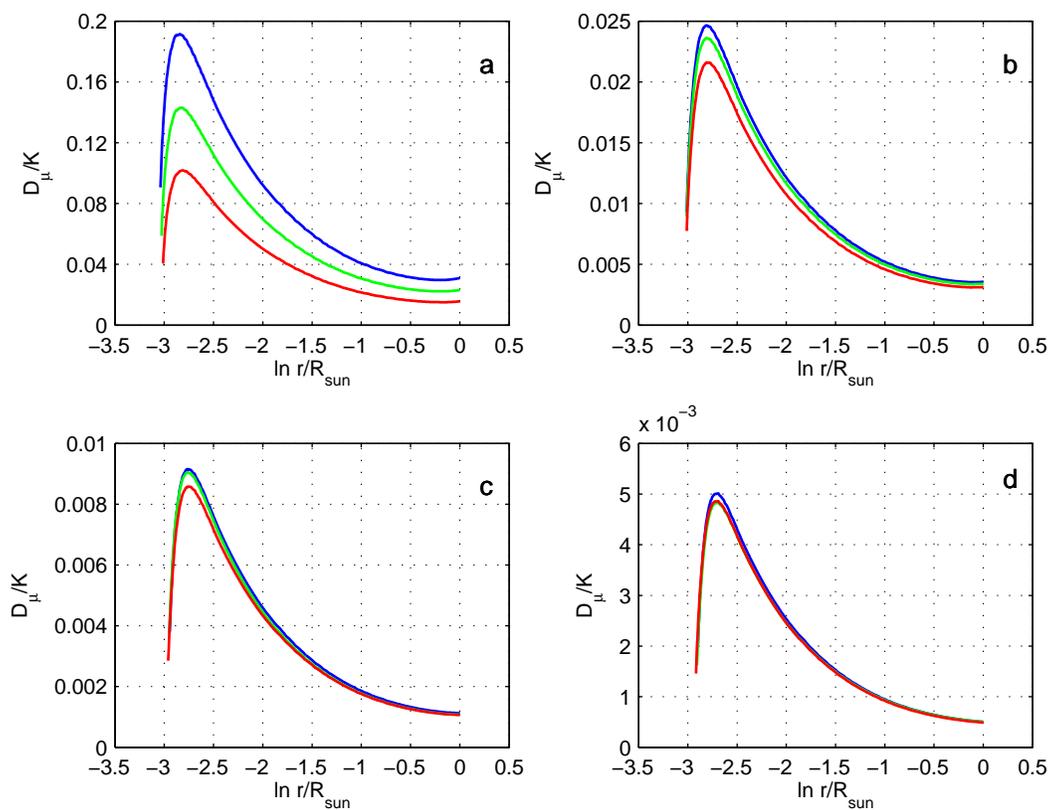}
\caption{Convergence of the diffusion coefficient profiles for the models presented by the same colors
         in the previous figure. Panels a, b, c, and d show how the results vary with time. Each panel is separated, in turn,  by
         a time interval that is approximately equal to 20\% of the total time for which the models
         have been evolved.
         }
\label{fig:f8}
\end{figure}

\begin{figure}
\epsffile [60 180 580 595] {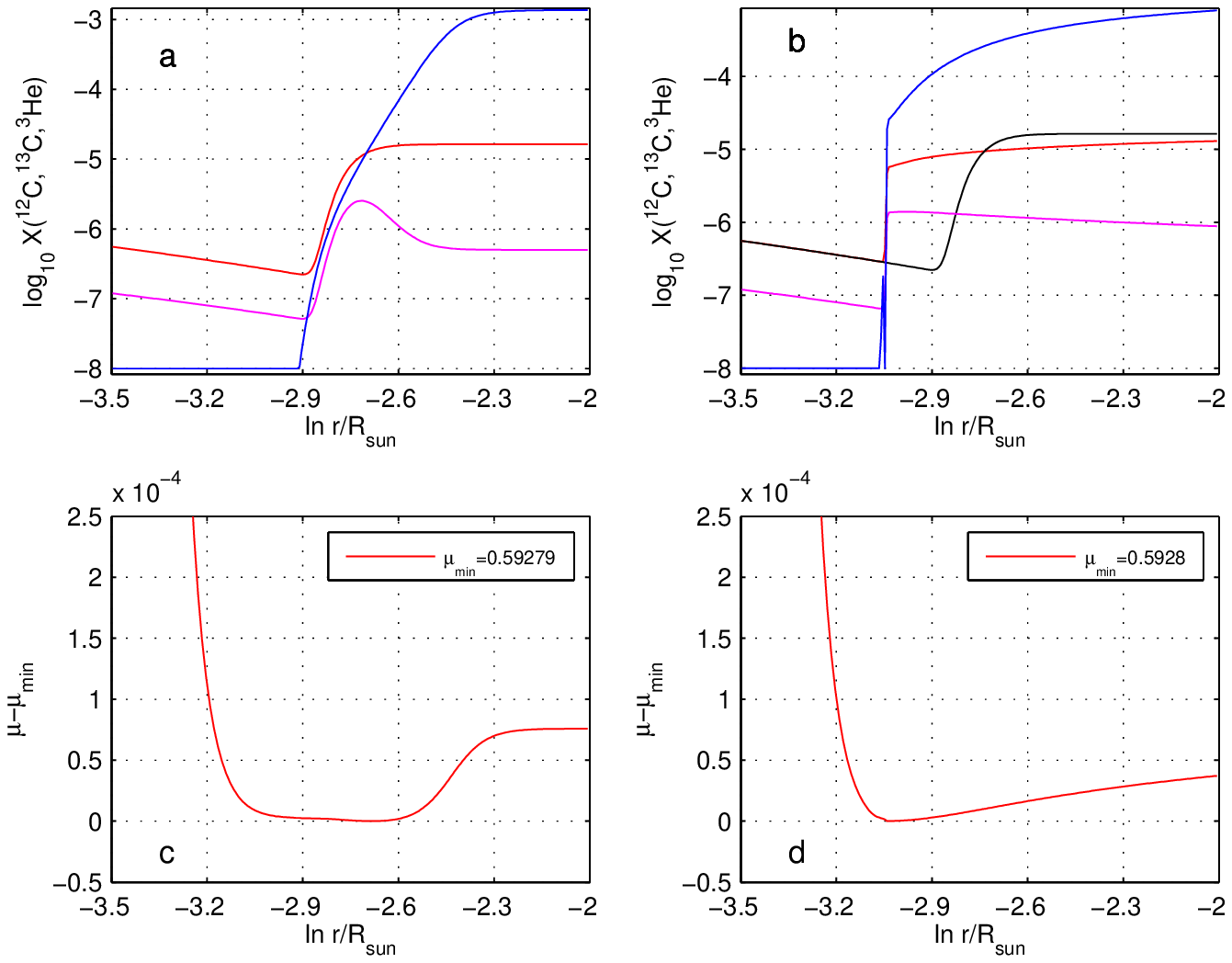}
\caption{Same as in Fig.~\ref{fig:f5} but for the bump luminosity model with $Z=0.0001$ and $M=0.85\,M_\odot$.
         }
\label{fig:f9}
\end{figure}

\begin{figure}
\epsffile [60 180 580 595] {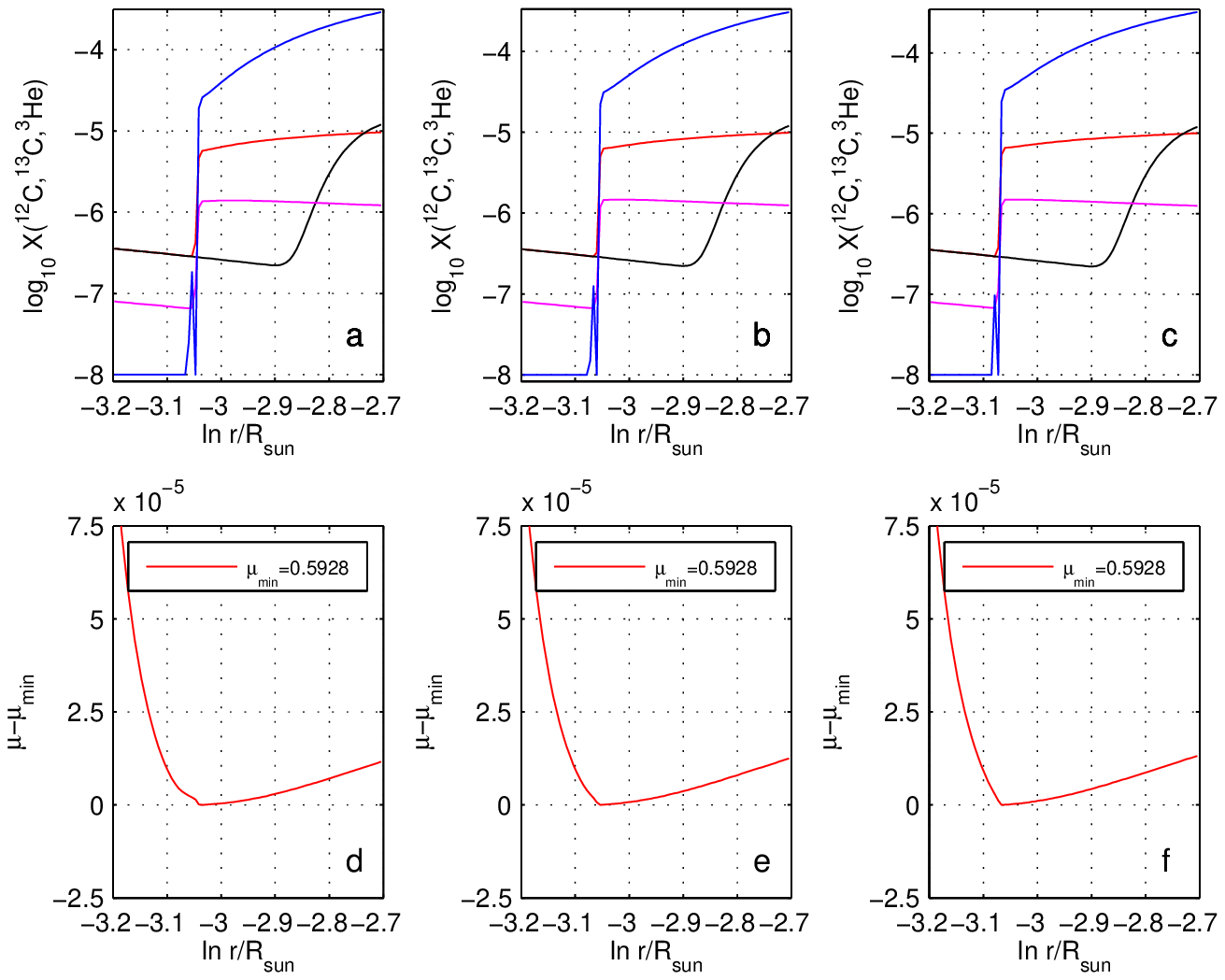}
\caption{Same as in Fig.~\ref{fig:f6} but for the bump luminosity model with $Z=0.0001$ and $M=0.85\,M_\odot$.
         }
\label{fig:f10}
\end{figure}

\begin{figure}
\epsffile [60 180 580 595] {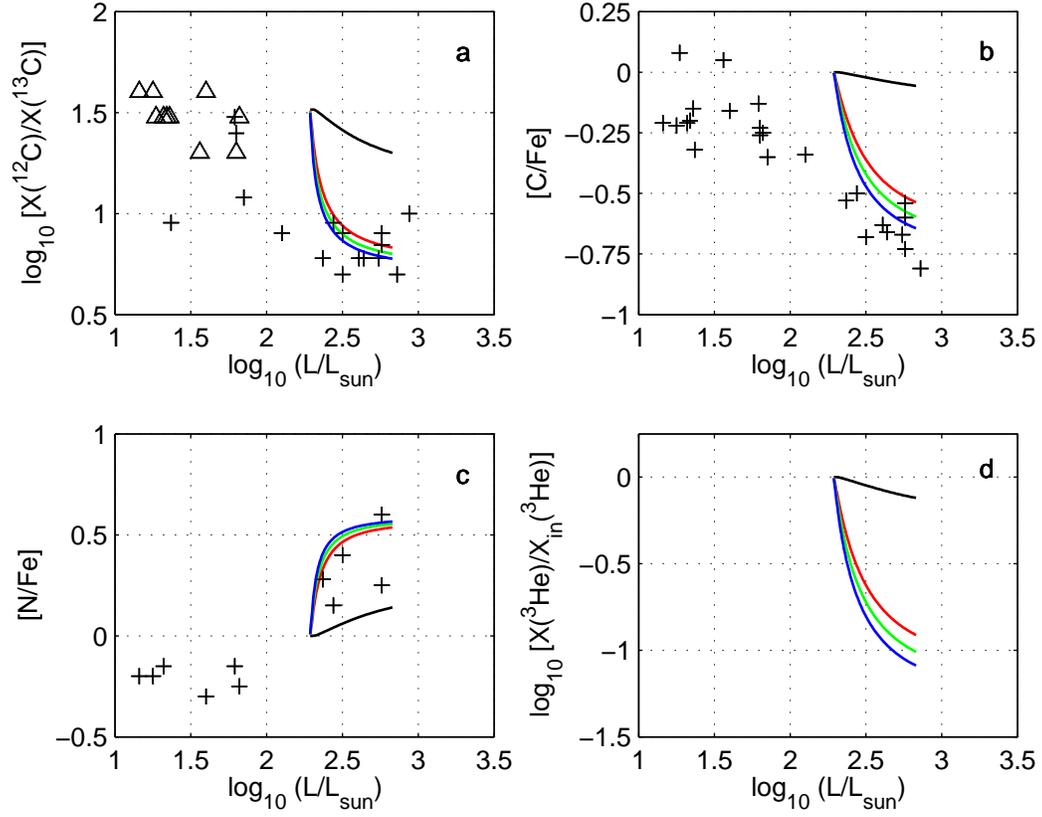}
\caption{Same as in Fig.~\ref{fig:f7}, except that the yellow curves are omitted, but for the RGB model
         with $Z=0.0001$ and $M=0.85\,M_\odot$.
         }
\label{fig:f11}
\end{figure}

\begin{figure}
\epsffile [60 180 580 595] {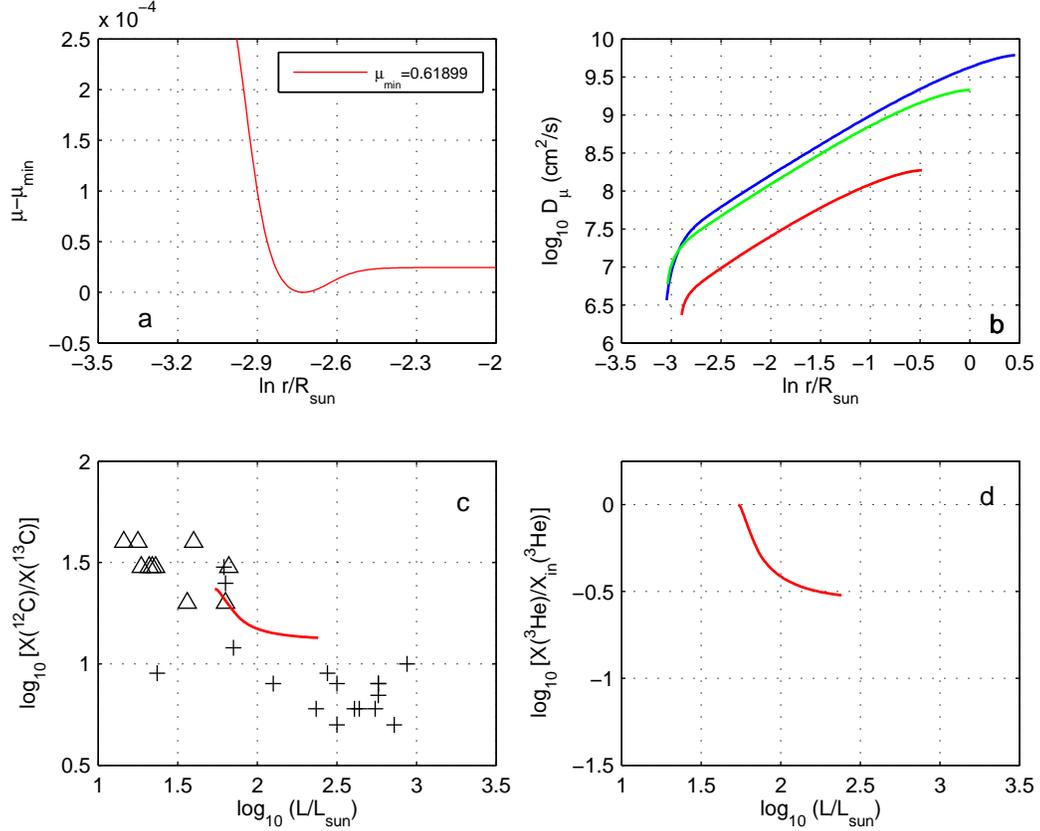}
\caption{The $\mu$ and $D_\mu$ profiles and the evolutionary changes of the $^{12}$C/$^{13}$C ratio and abundance of $^3$He computed
         for the RGB model with $Z=0.0188$ and $M=1.5\,M_\odot$ using the diffusion coefficient (\ref{eq:dmu}) with $a=7$ (red curves).
         The blue and green curves in panel b show the $D_\mu$ profiles from the bump luminosity models with $Z=0.0001$ and $Z=0.001$,
         respectively.
         }
\label{fig:f12}
\end{figure}

The $\mu$-profile in the vicinity of the HBS in our MESA bump luminosity model with $Z=0.0188$ and $M=1.5\,M_\odot$
is plotted in Fig.~\ref{fig:f12}. It is shown in the absence of mixing. It has two important differences, as compared to
the $\mu$-profiles in the low-metallicity models, by being shallower and narrower. The first property means that the $^3$He-driven thermohaline
mixing in the solar-metallicity RGB star should be much slower than that in low-metallicity red giants for the same finger aspect ratio. A comparison of
the coefficients $D_\mu$ for our three MESA models calculated using equation (\ref{eq:dmu}) with $a=7$ confirms
this conclusion (Fig.~\ref{fig:f12}). The second property means that the mixing in the solar-metallicity model
cannot penetrate as deep into the HBS as it did in the low-$Z$ models. As a result, it cannot dredge up carbon depleted material and, therefore,
does not affect the surface [C/Fe] abundance at all. The only visible abundance changes that it can cause are Li depletion,
and a modest reduction of the carbon isotopic ratio (Fig.~\ref{fig:f12}). It is interesting that
the linear theory (eq. \ref{eq:dmu}) with $a=7$ predicts that $^3$He-driven thermohaline mixing in the solar-metallicity
RGB star should lead to $^{12}$C/$^{13}\mbox{C}\approx 12$\,--\,$14$, which is very close to the abundance ratios measured
in upper RGB stars in the open cluster M\,67 (\citealt{gb91}), which has nearly the solar metallicity and an MS turn-off mass
of about $1.3\,M_\odot$ (e.g., \citealt{vs04}). This result shows that the $^3$He-driven thermohaline mixing modeled by
the diffusion coefficient (\ref{eq:dmu}) with the finger aspect ratio $a\ga 7$ can indeed reproduce properly the evolutionary
abundance changes observed in upper RGB stars of all metallicities, a conclusion similar to that obtained by \cite{chz07}.
Having said that, we emphasize that our 2D numerical simulations predict $a<1$; in which case, the model of the RGB thermohaline
mixing fails to interpret the observations (the black curves in Figs.~\ref{fig:f6} and \ref{fig:f10}). 

Li-rich giants may pose another problem for the explanation of the RGB extra-mixing by
the $^3$He-driven thermohaline convection. These stars have masses and metallicities close to those of our solar-metallicity
MESA model, and most of them are located near the bump luminosity (\citealt{chb00}). \cite{dh04} have shown that the anomalously large 
abundances of Li in these stars can be explained by the ``$^7$Be-transport'' mechanism (\citealt{cf71})
only if their radiative zones experience extra-mixing with a significantly enhanced diffusion coefficient 
(up to $10^{11}$\,cm$^2$\,s$^{-1}$). The blue curve in Fig.~\ref{fig:f12} corresponds to $a=7$ but it only reaches
values of the order of $D_\mu\sim 10^8$\,cm$^2$\,s$^{-1}$. To obtain the values necessary for efficient Li
production, we have to assume $a\approx 200$, which does not seem realistic. Hence, we would rather attribute the enhanced
extra-mixing in Li-rich giants to some alternative mixing process that replaces thermohaline convection
in these stars. However, we think this would be inconsistent because
it is difficult to understand why this alternative mixing process cannot be the universal one that operates
both in the Li-rich giants, which represent a few percent of all upper RGB stars, 
and in all other upper RGB stars in which its efficiency is reduced to $D_{\rm mix}\sim 10^8$\,--\,$10^9$\,cm$^2$\,s$^{-1}$
for some reason.
\cite{dea02} have reported that, in their sample of single K giants ``among rapid ($v\sin i\ga 8$\,km\,s$^{-1}$)
rotators, a very large proportion ($\sim$50\%) are Li-rich giants'' and that ``this proportion is in contrast with
a very low proportion ($\sim$2\%) of Li-rich stars among the much more common slowly rotating K-giants''.
This correlation of the RGB extra-mixing enhancement (needed for the Li enrichment) with the rapid rotation has been used
by \cite{dh04} to speculate that the RGB extra-mixing is actually driven by rotation and that the Li-rich giants
had been spun up as a result of their engulfing of massive planets. Note that, although \cite{pea06} have claimed that
the RGB extra-mixing cannot be associated with a pure rotational mechanism, there is still a possibility that
an interaction of rotation and large-scale magnetic fields can drive the mixing (\citealt{bea07,dea09}).
The only scenario that could explain this correlation
in the model with thermohaline mixing would be to assume that the Li-rich giants swallowed low-mass MS companions that are enriched
in $^3$He. This could explain the rapid rotation (a deposit of orbital angular momentum) and probably
the speed-up of extra-mixing by the increased amount of $^3$He that was supplied externally. However, from
our point of view, this scenario appears to be too complicated, and therefore highly improbable.

Finally, a comparison of panels d in Figs.~\ref{fig:f6}, \ref{fig:f10}, and \ref{fig:f12}
leads to the conclusion that, whereas $^3$He gets strongly depleted in the low-metallicity RGB stars, its envelope abundance
is reduced by a factor of only a few in the solar-metallicity model. On the one hand, this can be used as an argument against $^3$He-driven thermohaline
convection as the mechanism of RGB extra-mixing if the presence of similar extra-mixing process in low-metallicity
AGB stars with masses $M\la 1\,M_\odot$ surmised by \cite{mea06} and \cite{lea08} will by confirmed by other observations.
On the other hand, if the observational abundance anomalies, in particular those in meteorites (e.g., \citealt{nbw03}),
suggest the operation of an RGB-like extra-mixing only in the population I AGB stars with masses $M\ga 1\,M_\odot$
(\citealt{bea10}), then a more detailed analysis of the consequences of modest $^3$He depletion in the solar-metallicity RGB model
for the following AGB thermohaline mixing driven by $^3$He burning is still needed, especially given the arguments against
this hypothesis presented by \cite{kea10}. We will provide such the analysis in our forthcoming paper.

\section{Discussion and Conclusions}
\label{sec:concl}

In this work, we have come to the following two conclusions that happen to contradict one another. On the one hand, the linear stability analysis of
the Boussinesq equations (\ref{eq:bouss1}\,--\,\ref{eq:bouss3}) with parameters set up to describe the growth of
salt fingers driven by $^3$He burning in the vicinity of the HBS in a low-mass RGB star above the bump luminosity leads to
the thermohaline diffusion coefficient (\ref{eq:dmu}).  It models surprisingly well the observed evolutionary
changes of the surface chemical composition in upper RGB stars of different metallicities, provided that
we employ the empirically constrained finger aspect ratio $a_{\rm obs}\ga 7$. In other words, we arrive at a solution
very similar to that proposed by \cite{chz07}, namely that the effects of the RGB extra-mixing can be reproduced in
stellar evolutionary computations if the simple diffusion coefficient
\bea
D_\mu = 2\pi^2a_{\rm obs}^2\,\frac{\nabla_\mu}{\nabla_{\rm rad}-\nabla_{\rm ad}}\,K,
\label{eq:dmuobs}
\eea
where $a_{\rm obs} \ga 7$, is used. This large value of $a$ is close to the one advocated by \cite{u72}.
On the other hand, our 2D numerical simulations of thermohaline convection for the same RGB parameter set (Table~\ref{tab:tab1}) have shown that
the effective finger aspect ratio in expression (\ref{eq:dmu}) does not exceed a value of $a_{\rm eff}\approx 0.5$.
Interestingly, this small value of $a$ coincides with the one estimated by \cite{kea80}.
Because of the dependence of $D_\mu$ on the square of $a$, the difference between the two diffusion coefficients
calculated with $a=a_{\rm obs}$ and $a=a_{\rm eff}$ exceeds two orders of magnitude. It is highly unlikely
that our 2D numerical simulations have underestimated $D_\mu$ by this much. Therefore, we are inclined to conclude
that RGB extra-mixing has nothing to do with salt-fingering transport and that alternative mixing mechanisms
are worth investigating.

There is a clear physical explanation as to why salt-fingering leads to the turbulent convection in the RGB case, while
it produces vertically elongated quasi-laminar structures (salt fingers) in the oceanic case (e.g., \citealt{sh95,r10}).
The main reason for the different outcome is the large difference in the Prandtl number (Table~\ref{tab:tab1}).
The low Prandtl number (lower viscosity compared to heat diffusivity) in the RGB case favours the development of
secondary instabilities, predominantly the one associated with the mean field effects of vertical shear, which
do not allow the salt fingers to grow longer than their diameters in the vertical direction. We have demonstrated that
the artificial increase of viscosity does stabilize the growth of salt fingers, as expected. In real RGB stars,
such an increase could be associated with turbulence generated by rotation-driven instabilities, in which case
the turbulent viscosity $\nu_{\rm t}$ may exceed the microscopic (molecular plus radiative) viscosity $\nu$ by
several orders of magnitude (e.g., \citealt{z92,pea06}). However, it is difficult to imagine how the turbulence can
affect only the viscosity without enhancing chemical mixing at the same time. Hence, we also have to replace
$\nu_{\rm mol}$ by $\nu_{\rm t}$ in the expression in the parenthesis in the thermohaline diffusion
coefficient (\ref{eq:dmu}) that has been omitted in (\ref{eq:dmuobs}) for the sake of simplicity, because it can be neglected
in the absence of turbulence.
This replacement counterbalances the stabilizing effect of higher turbulent viscosity because turbulent mixing, especially
when it prevails in the horizontal direction (\citealt{z92}), facilitates the smoothing out of
the mean molecular weight contrast between the rising and sinking fingers, thus weakening the buoyancy force.
In connection with this, it is important to note that the hypothesis that RGB extra-mixing is only partially executed by
the $^3$He-driven thermohaline convection and that some other mixing mechanisms of rotational or magnetic origin
are assisting it (e.g., \citealt{cl10}), does not seem to be plausible. Indeed, we know from the comparison of
simple mixing models with the observations that extra-mixing in the majority of low-mass upper RGB stars needs
a diffusion coefficient of the order of $D_{\rm mix} \approx 0.02\,K$ (\citealt{dp08b}). If this mixing is associated with a mechanism different
from that of the salt-fingering convection then we have to use the ratio $D_{\rm mix}/K\approx 0.02$ instead of
$\nu_{\rm mol}/K$ in the parenthesis in the expression (\ref{eq:dmu}). Given that $R_\rho = (\nabla_{\rm rad}-\nabla_{\rm ad})/\nabla_\mu \ga 1000$
in the radiative zones of RGB stars, the expression in the parenthesis becomes negative, which means that the density stratification
is now stable against the double-diffusive instability. Consequently, there are only two possibilities: RGB extra-mixing is either
entirely executed by the $^3$He-driven thermohaline convection or it is the result of an entirely different mechanism.
To answer this question with certainty, our next step is to carry out 3D numerical simulations of thermohaline convection for the RGB case.

\acknowledgements
The author is grateful to Don VandenBerg who has supported this work through his Discovery Grant
from Natural Sciences and Engineering Research Council of Canada. The author also appreciates
discussions with Chris Garrett, Falk Herwig, Eric Kunze, and Bill Merryfield that have stimulated this work.
Special thanks go to Bill Merryfield for letting the author use his computer code designed for 2D numerical simulations of
salt-fingering convection and for making available his unpublished manuscript.


\clearpage
\begin{deluxetable}{lllll}
\tablecolumns{5}
\tabletypesize{\footnotesize}
\tablecaption{Correspondence Between Salt-Fingering Parameters}
\tablewidth{0pt}
\tablehead{
  \colhead{} & \multicolumn{2}{c}{Oceanic Case} &
  \multicolumn{2}{c}{RGB Case} \\
  \cline{2-3} \cline{4-5} \\
  \colhead{Parameter} &
            \colhead{Notation} & \colhead{Value (cgs) } &
                      \colhead{Notation} & \colhead{Value (cgs) } }
                      \startdata
                      Viscosity & $\nu$ & $10^{-2}$ & $\nu$ & $4\times 10^2$ \\
                      Thermal Diffusivity & $k_T$ & $1.4\times 10^{-3}$ & $K$ & $10^8$ \\
                      Haline Diffusivity & $k_S$ & $1.1\times 10^{-5}$ & $\nu_{\rm mol}$ & $2\times 10^2$ \\
                      Gravitational Acceleration & $g$ & $9.8\times 10^2$ & $g$ & $10^6$ \\
                      Thermal Expansion & $\alpha$ & $2\times 10^{-4}$ & $\alpha\approx T^{-1}$ & $10^{-7}$ \\
                      Haline Contraction & $\beta$ & $7.5\times 10^{-4}\ {^0/_{00}}^{-1}$ & $\beta\approx\mu^{-1}$ & $1.7$ \\
                      Temperature Gradient & $\frac{\partial T_0}{\partial z}$ & $3\times 10^{-3}$ & $\frac{\partial T_0}{\partial z} - 
                         \left(\frac{\partial T}{\partial z}\right)_{\rm ad}$ & $2\times 10^{-3}$ \\
                      Density Ratio & $R_\rho$ & $1.6$ & $R_\rho = (\nabla - \nabla_{\rm ad})/\nabla_\mu$ & $1.7\times 10^3$ \\
                      Prandtl Number ($\nu/k_T$) & $Pr$ & $7$ & $Pr$ & $4\times 10^{-6}$ \\
                      Inverse Lewis Number ($k_S/k_T$) & $\tau$ & $8\times 10^{-3}$ & $\tau$ & $2\times 10^{-6}$ \\
                      \enddata
\label{tab:tab1}
\end{deluxetable}


\end{document}